\pdfoutput=1

\documentclass[iop,apj,a4paper,numberedappendix,appendixfloats]{emulateapj}
\usepackage{apjfonts,amsmath,amssymb,epsf,lscape}

\begin{document}

\shortauthors{Kobayashi et al.}
\shorttitle{HST/ACS morphology of Ly$\alpha$ emitters at $z = 4.86$}

\slugcomment{Received 2015 February 10; accepted 2016 January 19; published 2016 xxx}

\title{MORPHOLOGICAL PROPERTIES OF LYMAN $\alpha$ EMITTERS AT REDSHIFT
4.86 IN THE COSMOS FIELD: CLUMPY STAR FORMATION OR
MERGER?\altaffilmark{*}}

\author{
Masakazu~A.~R.~Kobayashi\altaffilmark{1},
Katsuhiro~L.~Murata\altaffilmark{2},
Anton~M.~Koekemoer\altaffilmark{3},
Takashi~Murayama\altaffilmark{4},
Yoshiaki~Taniguchi\altaffilmark{1},
Masaru~Kajisawa\altaffilmark{1,5},
Yasuhiro~Shioya\altaffilmark{1},
Nick~Z.~Scoville\altaffilmark{6},
Tohru~Nagao\altaffilmark{1}, and
Peter~L.~Capak\altaffilmark{6,7}
}

\email{kobayashi@cosmos.phys.sci.ehime-u.ac.jp}

\altaffiltext{*}{Based on observations with NASA/ESA \textit{Hubble
Space Telescope}, obtained at the Space Telescope Science Institute,
which is operated by AURA, Inc., under NASA contract NAS 5-26555; and
also based on data collected at Subaru Telescope, which is operated by
the National Astronomical Observatory of Japan.}
\altaffiltext{1}{Research Center for Space and Cosmic Evolution, Ehime
University, Bunkyo-cho 2-5, Matsuyama 790-8577, Japan}
\altaffiltext{2}{Department of Physics, Nagoya University, Furo-cho,
Chikusa-ku, Nagoya 464-8602}
\altaffiltext{3}{Space Telescope Science Institute, 3700 San Martin
Drive, Baltimore, MD 21218}
\altaffiltext{4}{Astronomical Institute, Graduate School of Science,
Tohoku University, Aramaki, Aoba, Sendai 980-8578, Japan}
\altaffiltext{5}{Graduate School of Science and Engineering, Ehime
University, Bunkyo-cho, Matsuyama 790-8577, Japan}
\altaffiltext{6}{Department of Astronomy, MS 105-24, California
Institute of Technology, Pasadena, CA 91125}
\altaffiltext{7}{Spitzer Science Center, California Institute of
Technology, Pasadena, CA 91125}

\begin{abstract}

 We investigate morphological properties of 61 Ly$\alpha$ emitters
 (LAEs) at $z=4.86$ identified in the COSMOS field, based on
 \textit{Hubble Space Telescope} Advanced Camera for Surveys (ACS)
 imaging data in the F814W-band.  Out of the 61 LAEs, we find the ACS
 counterparts for the 54 LAEs.  Eight LAEs show double-component
 structures with a mean projected separation of 0\farcs63
 ($\sim4.0$~kpc at $z=4.86$).  Considering the faintness of these ACS
 sources, we carefully evaluate their morphological properties, that
 is, size and ellipticity.  While some of them are compact and
 indistinguishable from the PSF half-light radius of 0\farcs07
 ($\sim0.45$~kpc), the others are clearly larger than the PSF size and
 spatially extended up to 0\farcs3 ($\sim1.9$~kpc).  We find that the
 ACS sources show a positive correlation between ellipticity and size
 and that the ACS sources with large size and round shape are absent.
 Our Monte Carlo simulation suggests that the correlation can be
 explained by (1) the deformation effects via PSF broadening and shot
 noise or (2) the source blending in which two or more sources with
 small separation are blended in our ACS image and detected as a
 single elongated source.  Therefore, the 46 single-component LAEs
 could contain the sources which consist of double (or multiple)
 components with small spatial separation (i.e., $\lesssim0\farcs3$
 or 1.9~kpc).  Further observation with high angular resolution at
 longer wavelengths (e.g., rest-frame wavelengths of
 $\gtrsim4000$~{\AA}) is inevitable to decipher which interpretation
 is adequate for our LAE sample.

\end{abstract}


 \section{INTRODUCTION}

 In the standard picture of structure formation, within the framework
 of Cold Dark Matter (CDM) models, small subgalactic clumps are formed
 first in CDM halos.  Such building blocks of normal galaxies in local
 universe grow hierarchically into more massive galaxies through
 galaxy mergers and subsequent star formation.  Ly$\alpha$ emitters
 (LAEs) at high-$z$ universe are considered to be building blocks
 because of their small stellar masses, young ages, and low
 metallicities inferred from their broadband spectral energy
 distributions (e.g., Chary et al. 2005; Gawiser et al. 2006; Nilsson
 et al. 2007, 2009; Finkelstein et al. 2008; Ono et al. 2010a,~b; Yuma
 et al. 2010; Acquaviva et al. 2011; Guaita et al. 2011; Vargas et
 al. 2014).  Since they are important population as a probe of galaxy
 formation in the young universe as well as a probe of cosmic
 reionization, much effort has been paid to search them (e.g., Cowie
 \& Hu 1998; Rhoads et al. 2000; Ouchi et al. 2005, 2008, 2010;
 Taniguchi et al. 2005; Shimasaku et al. 2006; Gronwall et al. 2007;
 Murayama et al. 2007; Shioya et al. 2009; Kashikawa et al. 2011).
 The redshift of the most distant LAE has now reached beyond $z = 7$
 (Ono et al. 2012; Shibuya et al. 2012; Finkelstein et al. 2013), at
 which cosmic reionization has not been completed yet.

 However, it is still unclear in what physical conditions a galaxy is
 observed as an LAE, which has intense Ly$\alpha$ emission.  This is
 mainly because Ly$\alpha$ is a resonance line of neutral hydrogen;
 that is, mean free path of Ly$\alpha$ photon in interstellar medium
 (ISM) is significantly short and hence it experiences enormous number
 of scattering by neutral hydrogen before escaping from its host
 galaxy.  The multiple scattering makes Ly$\alpha$ extremely
 vulnerable to dust attenuation.  This is consistent with the
 observational results for the LAEs in both nearby and high-$z$
 universe which have revealed that the Ly$\alpha$ escape fraction
 depends clearly on dust extinction, although the escape fraction does
 not follow the expected one for a simple attenuation (Atek et
 al. 2009, 2014; Kornei et al. 2010; Hayes et al. 2011, 2014).
 Theoretical studies have also been executed, in which Ly$\alpha$
 radiative transfer code is coupled with cosmological numerical
 simulation in order to examine the Ly$\alpha$ escape fraction in
 realistic ISM condition for high-$z$ LAE (e.g., Laursen \&
 Sommer-Larsen 2007; Laursen et al. 2009a,~b; Zheng et al. 2010;
 Yajima et al. 2012a,~b).  These theoretical studies predict that ISM
 clumpiness and morphology have a strong impact on Ly$\alpha$ escape
 fraction and that clumpy and dusty ISM is favored for Ly$\alpha$ to
 escape (Yajima et al. 2012b; Laursen et al. 2013; Duval et al. 2014;
 Gronke \& Dijkstra 2014).  Moreover, such clumpy and dusty ISM is
 also found to be favored to reproduce the observed statistical
 properties of LAEs (Kobayashi et al. 2007, 2010).

 In such context, observational studies for the size and morphology of
 high-$z$ LAEs have been widely conducted by using the Advanced Camera
 for Surveys (ACS) on-board the \textit{Hubble Space Telescope}
 (\textit{HST}) because these properties give us insights on how LAEs
 are assembled and how their intense star formation events are
 triggered (e.g., Stanway et al. 2004; Rhoads et al. 2005; Venemans et
 al. 2005; Pirzkal et al. 2007; Overzier et al. 2008; Bond et
 al. 2009, 2012; Taniguchi et al. 2009; Vanzella et al. 2009;
 Finkelstein et al. 2011; Law et al. 2012; Malhotra et al. 2012;
 Mawatari et al. 2012; Chonis et al. 2013; Jiang et al. 2013; Hagen et
 al. 2014; Shibuya et al. 2014).  It has been found that most of the
 high-$z$ LAEs have small sizes of 0\farcs 1--0\farcs 2 in rest-frame
 ultraviolet (UV) continuum, which remain almost constant in the
 redshift range of $z \sim 2$--6 (Malhotra et al. 2012; Hagen et
 al. 2014).  This is against the hypothesis that LAE is simply a
 subset of Lyman-break galaxy (LBG) population, which present a clear
 redshift evolution of size in rest-frame UV continuum (e.g., Ono et
 al. 2013).

 In this paper, we examine the morphological properties of the 61 LAEs
 at $z = 4.86$ selected by Shioya et al. (2009; hereafter S09) in the
 Cosmic Evolution Survey (COSMOS) field (Scoville et al. 2007a),
 providing one of the largest samples of LAEs in a large contiguous
 field.  Since F814W-band imaging taken with the \textit{HST}/ACS is
 available for the COSMOS field (Scoville et al. 2007b; Koekemoer et
 al. 2007), the sizes and morphologies of the LAEs in the COSMOS field
 can be investigated in detail.  In this paper, we present our
 detailed analysis of ACS images of the LAE sample of S09.

 We use a standard cosmology with $\Omega_M = 0.3$, $\Omega_\Lambda =
 0.7$, and $H_0 = 70~\mathrm{km~s^{-1}~Mpc^{-1}}$.  Under the adopted
 cosmological parameters, the angular scale of $1^{\prime\prime}$
 corresponds to the physical scale of 6.37~kpc at $z = 4.86$.
 Throughout this paper, we use magnitudes in the AB system.

 \section{OBSERVATIONAL DATA AND ACS COUNTERPARTS OF
 LAEs}\label{sec:data}

\begin{figure}
  \epsscale{1.15}
  \plotone{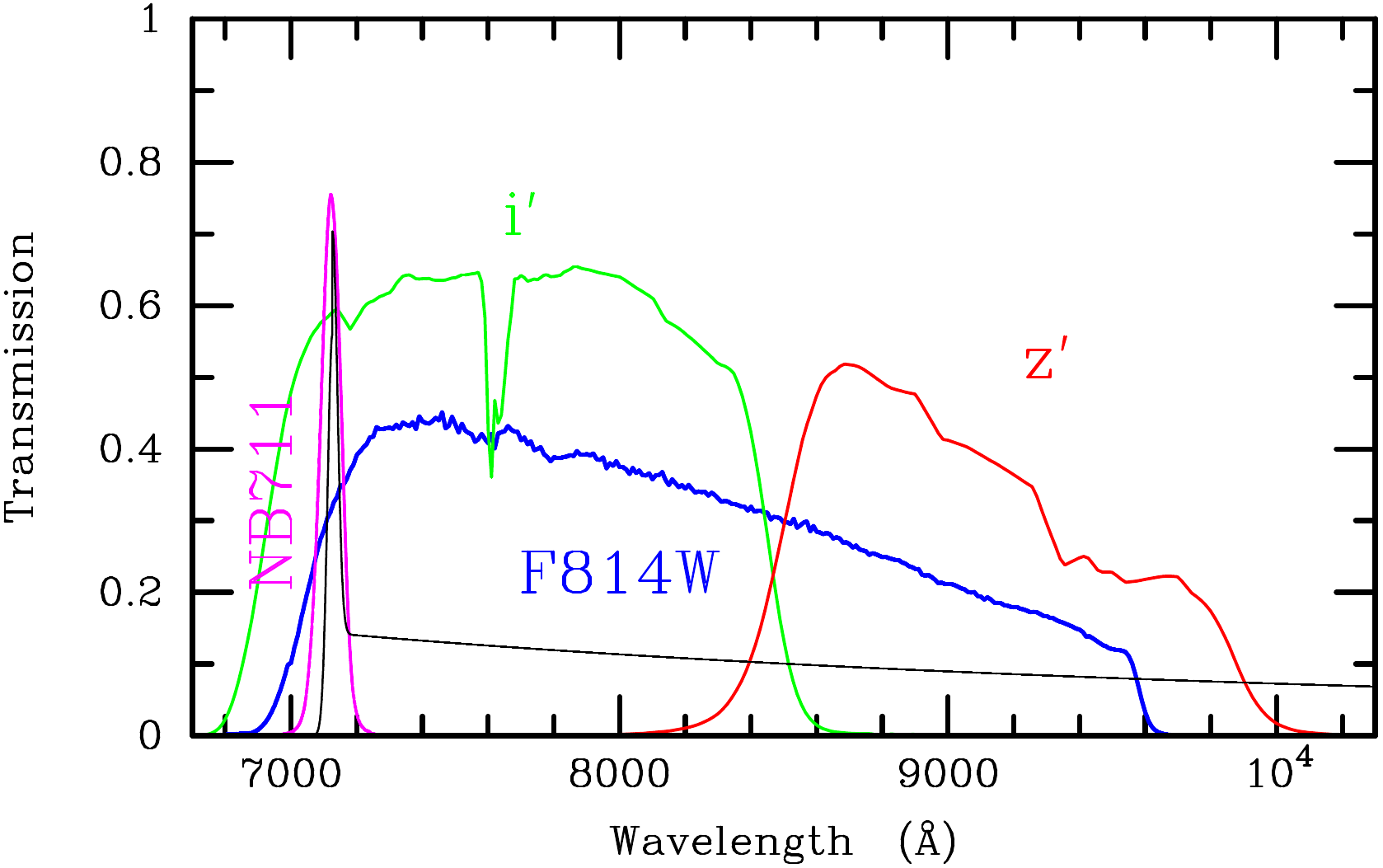}

  \caption{Transmission curves for the filters related to our
  analysis.  The blue curve represents the transmission curve for the
  \textit{HST}/ACS F814W-band, while the magenta, green, and red
  curves are the transmissions for the Subaru/Suprime-Cam NB711--,
  $i^\prime$--, and $z^\prime$--bands, respectively.  The effects of
  the CCD sensitivity, the atmospheric transmission, and the
  transmission of the telescope and the instrument are taken into
  account for each transmission curve.  Model spectrum of a LAE at $z
  = 4.86$ with a rest-frame Ly$\alpha$ equivalent width
  ($\mathrm{EW_0}$) of 30~{\AA} is also plotted by black
  curve. \label{fig:filters}}

 \end{figure}

 In S09, 79 LAE candidates at $ 4.83 < z < 4.89$ have been carefully
 selected from optical imaging with the narrow-band filter, NB711
 ($\lambda_\mathrm{c} = 7126$~{\AA}, $\Delta \lambda = 73$~{\AA}; see
 Figure~\ref{fig:filters}), and broad-band filters from $B$ to
 $z^\prime$ taken for the entire $1.95~\mathrm{deg^2}$ area of the
 COSMOS field using the Suprime-Cam (Miyazaki et al. 2002) on the
 Subaru Telescope (Kaifu et al. 2000; Iye et al. 2004).  Details of
 the Subaru observations and data processing are described by
 Taniguchi et al. (2007) and Capak et al. (2007).  Among 79 LAEs, 13
 LAEs have spectroscopic information and all of them are confirmed as
 $z \approx 4.86$ (P. Capak et al. 2015, in preparation), verifying
 the effectiveness of our selection method\footnote{Although a
 follow-up spectroscopy has also been performed for 5 additional LAEs,
 their spectroscopic redshifts have not been determined because of low
 data quality (see Table~\ref{tab:z4p9LAE}).}.

 \begin{figure}
  \epsscale{1.15}
  \plotone{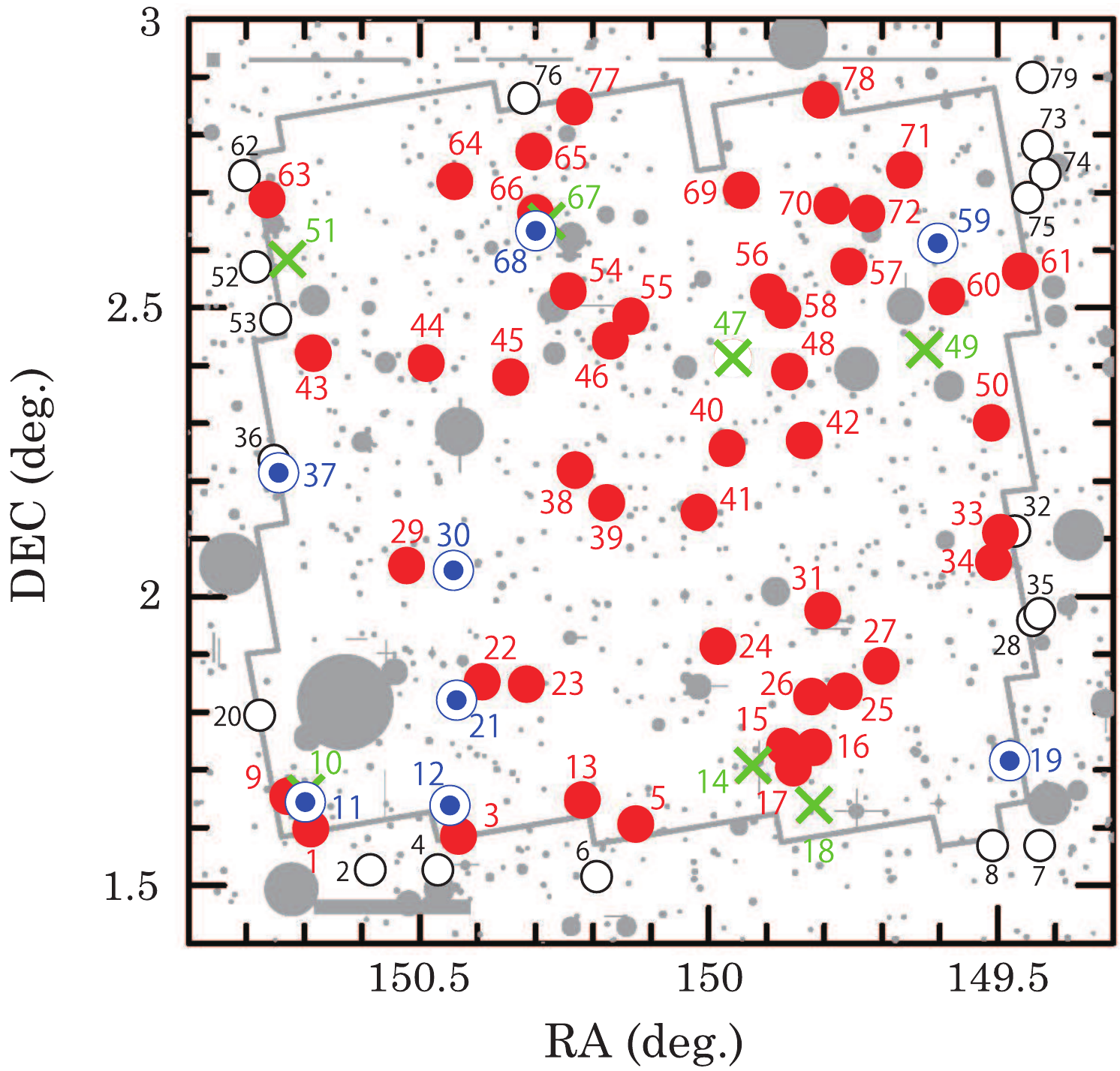}

  \caption{Spatial distribution of our sample of 79 LAEs at $z = 4.86$
  selected in S09.  In the whole COSMOS field of
  $1.95~\mathrm{deg}^2$, the \textit{HST}/ACS images are available for
  $1.64~\mathrm{deg}^2$ indicated by the solid gray line.  The gray
  shaded regions represent the areas masked out for detection.  The 18
  LAEs outside the \textit{HST}/ACS field are shown by black open
  circles.  Among the remaining 61 LAEs in the \textit{HST}/ACS field,
  7 LAEs undetected in the ACS images are represented by green crosses
  and 46 (8) LAEs with single (double) component(s) are shown by red
  filled (blue double) circles.  The ID \# in S09 is labeled for
  reference.\label{fig:map}}

 \end{figure}
 \begin{deluxetable*}{lll}
  \tablecaption{SExtractor configuration for the HST/ACS F814W-band
  detection \label{table:SExParam}}
  \tablehead{
  \colhead{Parameter} & \colhead{Value} & \colhead{Comment}
  }
  \startdata
  DETECT\_THRESH   & 1.1 & Detection threshold in sigma\\
  DETECT\_MINAREA  & 25  & Minimum number of pixels above threshold\\
  FILTER\_NAME     & gauss\_3.0\_7x7.conv & Name of the filter for detection\\
  DEBLEND\_NTHRESH & 64  &  Number of deblending sub-thresholds\\
  DEBLEND\_MINCONT & 0.015 & Minimum contrast parameter for deblending\\
  PHOT\_AUTOPARAMS & 2.5, 0.5 & MAG\_AUTO parameters: Kron factor and minimum radius\\
  BACK\_SIZE       & 64 &  Background mesh size\\
  BACK\_FILTERSIZE &  3 &  Background filter size\\
  BACKPHOTO\_TYPE  & GLOBAL &  Photometry background subtraction type
  \enddata
 \end{deluxetable*}

 The \textit{HST}/ACS F814W-band data ($\lambda_\mathrm{c} =
 8333$~{\AA}, $\Delta \lambda = 2511$~{\AA}; see
 Figure~\ref{fig:filters}) is available for a part of the COSMOS
 field, $1.64~\mathrm{deg^2}$ ($\approx 84$\% of the COSMOS field), as
 shown in Figure~\ref{fig:map}.  In our analysis, we use the official
 COSMOS ACS image (Scoville et al. 2007b; Koekemoer et al. 2007),
 Version 2.0.  The ACS data were processed to $0\farcs
 03~\mathrm{pixel}^{-1}$ images.  We find that ACS imaging data are
 available for 61 LAEs out of 79 LAE samples selected by S09.  The
 remaining 18 LAEs are not covered by the ACS field or are on the edge
 of the ACS field.  Spatial distribution of all 79 LAEs in the COSMOS
 field is shown in Figure~\ref{fig:map}.  Our data analysis procedure
 for ACS data are similar to those in Taniguchi et al. (2009), in
 which the official COSMOS ACS image Version 1.3 with the pixel scale
 of $0\farcs 05~\mathrm{pixel}^{-1}$ was utilized.  The source
 detection of the LAEs in the \textit{HST}/ACS image was carried out
 with their weight map using SExtractor (Bertin \& Arnouts 1996).  The
 fundamental parameters of the SExtractor's configuration are shown in
 Table~\ref{table:SExParam}, which are also basically similar to those
 used in Taniguchi et al. (2009) and modified slightly for the
 $0\farcs 03~\mathrm{pixel}^{-1}$ images.  Note that these parameters
 are determined by the tradeoff between detecting fainter
 objects/components and avoiding noise effects such as the false
 detection or the noise confusion.

 \begin{deluxetable}{lllrr}
  \tablecaption{COSMOS $z = 4.86$ LAE Sample \label{table:Sample}}
  \tablewidth{0pt}
  \tablehead{
  \multicolumn{3}{c}{LAE Sample} & \colhead{Number of LAEs} &
  \colhead{Spectroscopic}\\
  \multicolumn{4}{c}{} & \colhead{Confirmation}
  }
  \startdata
  \multicolumn{3}{l}{In the ACS/F814W-band field}     & 61 & 13 \vspace{1mm}\\
  & \multicolumn{2}{l}{ACS/F814W-band detected}       & 54 & 12 \vspace{1mm}\\
  & & \hspace{4mm}Single component               & 46 & 10 \vspace{1mm}\\
  & & \hspace{4mm}Double component               &  8 & 2 \vspace{1mm}\\
  & \multicolumn{2}{l}{ACS/F814W-band undetected}     &  7 & 1 \vspace{1mm}\\
  \multicolumn{3}{l}{Out of the ACS/F814W-band field} & 18 & 0 \vspace{1mm}\\
  \cline{1-5}\vspace{-2mm}\\
  \multicolumn{3}{c}{Total} & 79 & 13
  \enddata
 \end{deluxetable}
 Among the 61 LAEs in the ACS field, we find the ACS counterparts of
 the 54 LAEs detected near the LAE positions defined in NB711--band
 images (i.e., separation of $\le 1^{\prime\prime}$).  Any sources are
 not detected near the LAE positions for the remaining seven LAEs.
 While most of the ACS counterparts consist of single component, eight
 LAEs among the 54 ACS-detected LAEs have double components in the ACS
 images within separation of $\le 1^{\prime\prime}$ from the LAE
 positions, providing the double-component LAE fraction,
 $f_\mathrm{double}$, of $8 / 54 = 14.8$\%.  The numbers of the total
 sample, both the ACS-detected and undetected LAEs, are summarized in
 Table~\ref{table:Sample}.

 Separations between each component in the 8 double-component LAEs are
 found to be 0\farcs 36--0\farcs 98 (the mean is 0\farcs 63).  For
 these double-component LAEs, the mean offset of the ACS centroids
 from the NB711--band centroids is found to be 0\farcs 39, which is
 larger than the NB711--band pixel scale of $0\farcs
 15~\mathrm{pixel}^{-1}$.  This is possibly because these double ACS
 components are unresolved in the NB711--band images taken by the
 Subaru telescope, in which the mean half-light radius of unsaturated
 stars is 0\farcs 25, and their NB711--band positions can be close to
 their flux-weighted centroid.  On the other hand, for the 46
 single-component LAEs, the mean offset between the ACS F814W- and
 NB711--band centroids is 0\farcs 16, which is comparable to the pixel
 scale of the NB711--band images.  We note that, in the following
 analysis, the double ACS components in each one of the 8
 double-component LAEs are treated as sub-components in a single
 object at first and the morphological properties of the objects are
 measured.  The results with each of the components in the double
 component systems treated separately as different objects with close
 angular separation are presented in Section~\ref{subsec:double}.

 As shown in Figure~\ref{fig:map}, the ACS-detected (filled and double
 circles) and ACS-undetected LAEs (crosses) seem to be distributed
 randomly in the whole ACS field.  Therefore, their distributions may
 not be affected by large-scale inhomogeneity of the ACS data quality
 (e.g., edges of the field).  It should be noted that the LAE \#20 is
 on the edge of the ACS field as shown in Figure~\ref{fig:map}.  While
 the LAE \#20 seems to have a double-component ACS source, we do not
 include it in our sample of the ACS-detected LAEs since the ACS data
 quality is highly doubtful.

 \begin{figure}
  \epsscale{1.1}
  \plotone{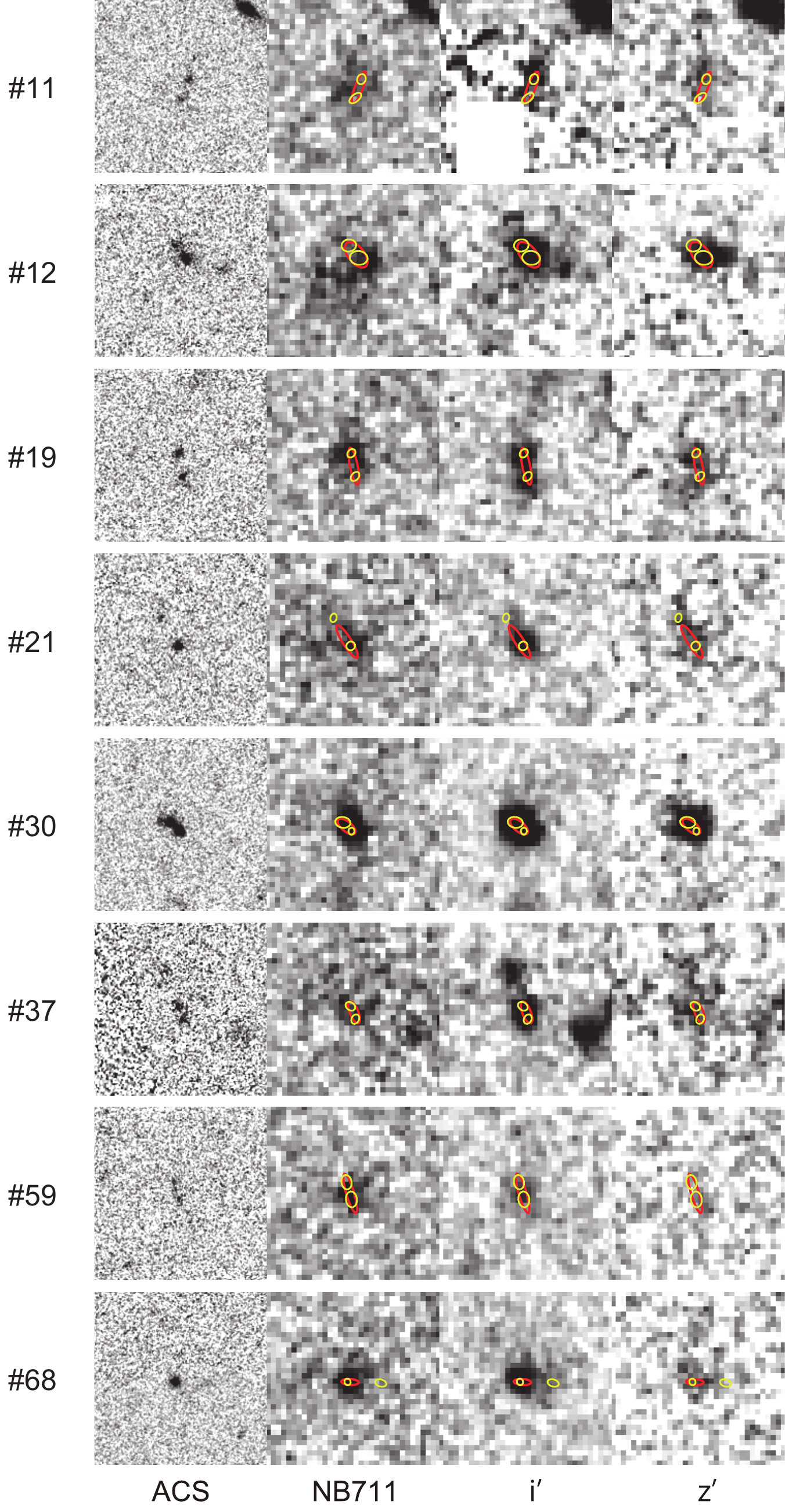}

  \caption{Thumbnails of 8 LAEs with double-component ACS F814W-band
  sources.  North is up and east is left.  Each panel has a size of
  $5^{\prime\prime} \times 5^{\prime\prime}$.  Red ellipses
  overplotted on the NB711-- and $z'$--band images are half light
  ellipses of the detected LAE counterparts in the ACS image, while
  yellow ellipses are those of the individual components in each LAE
  counterpart detected by SExtractor. \label{fig:ThumbnailsDouble}}

 \end{figure}
 \begin{figure*}
  \epsscale{1.1}
  \plotone{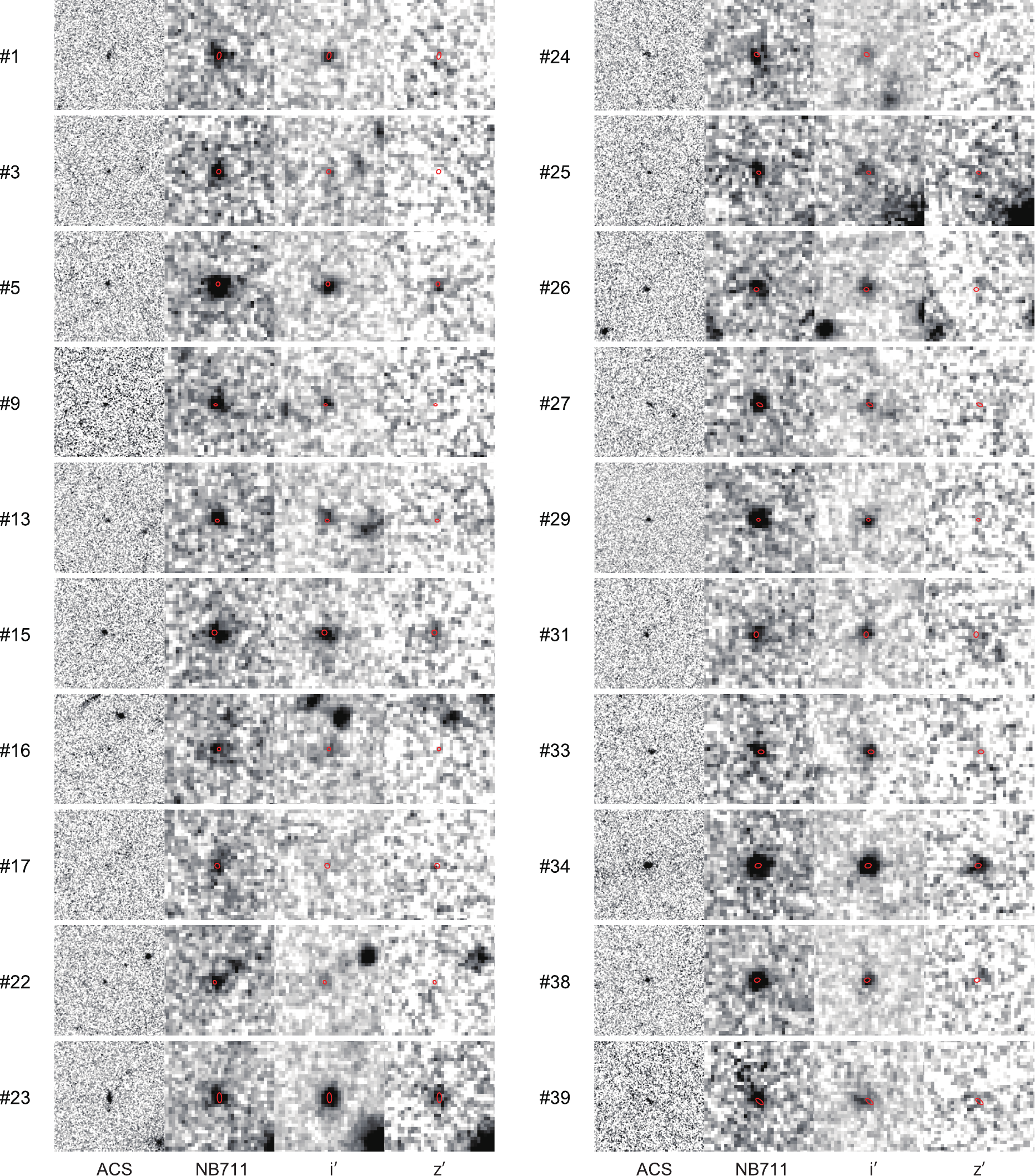}
  \caption{Same as Figure~\ref{fig:ThumbnailsDouble} but for 46 LAEs
  with single component ACS source. \label{fig:ThumbnailsSingle}}
 \end{figure*}
 \setcounter{figure}{3}
 \begin{figure*}
  \epsscale{1.1}
  \plotone{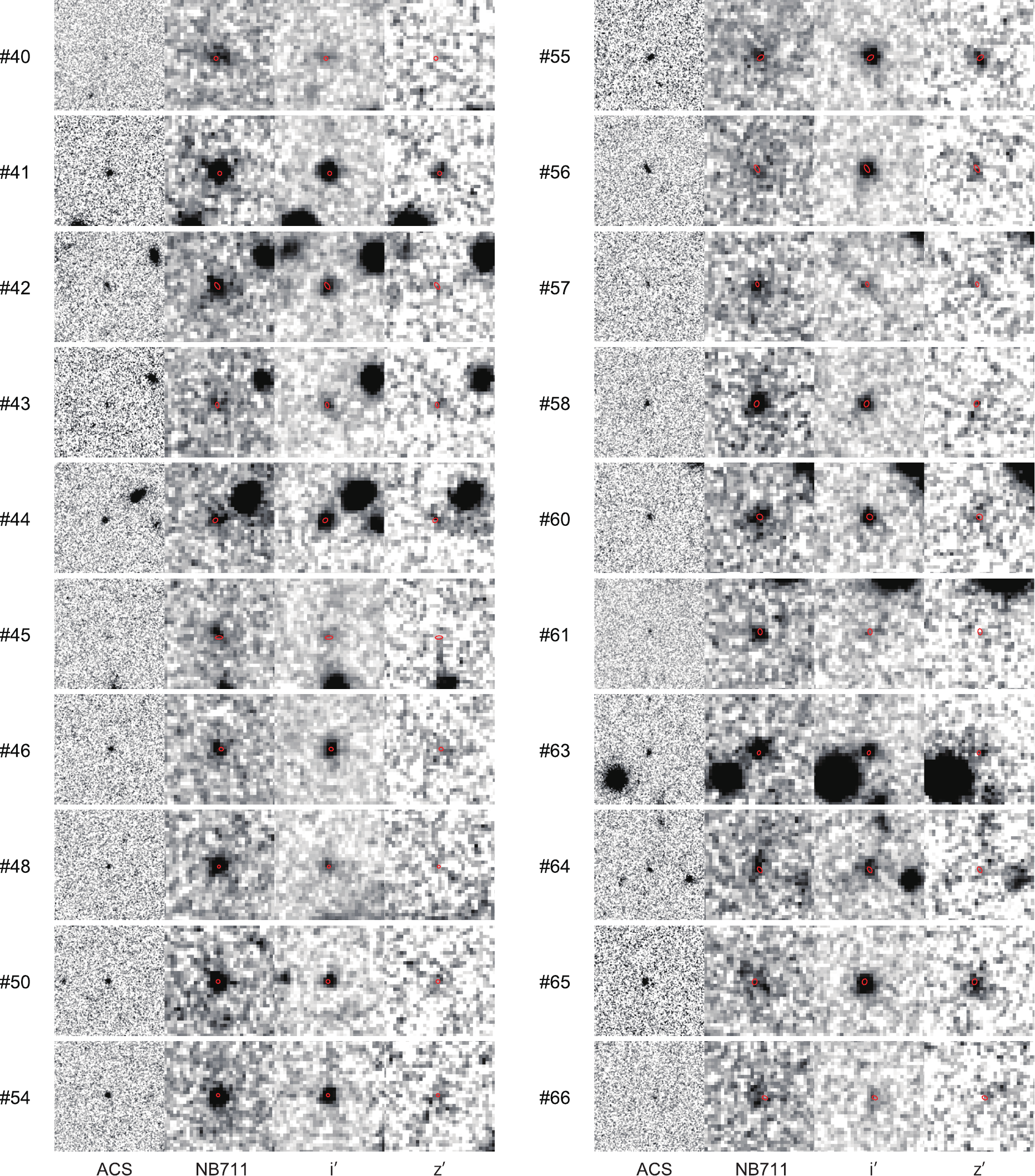}
  \caption{(Continued.)}
 \end{figure*}
 \setcounter{figure}{3}
 \begin{figure}
  \epsscale{1.1}
  \plotone{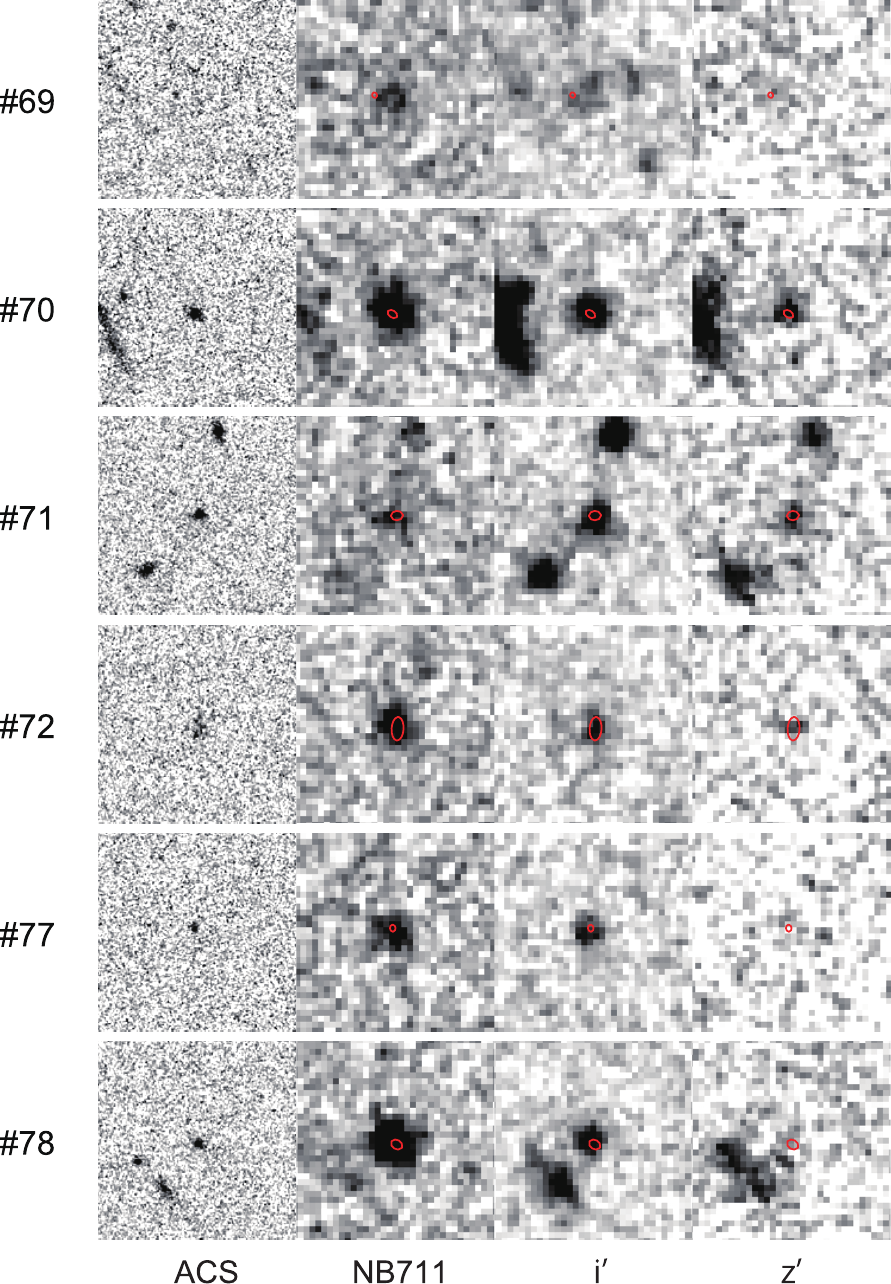}
  \caption{(Continued.)}
 \end{figure}
 \begin{figure}
  \epsscale{1.1}
  \plotone{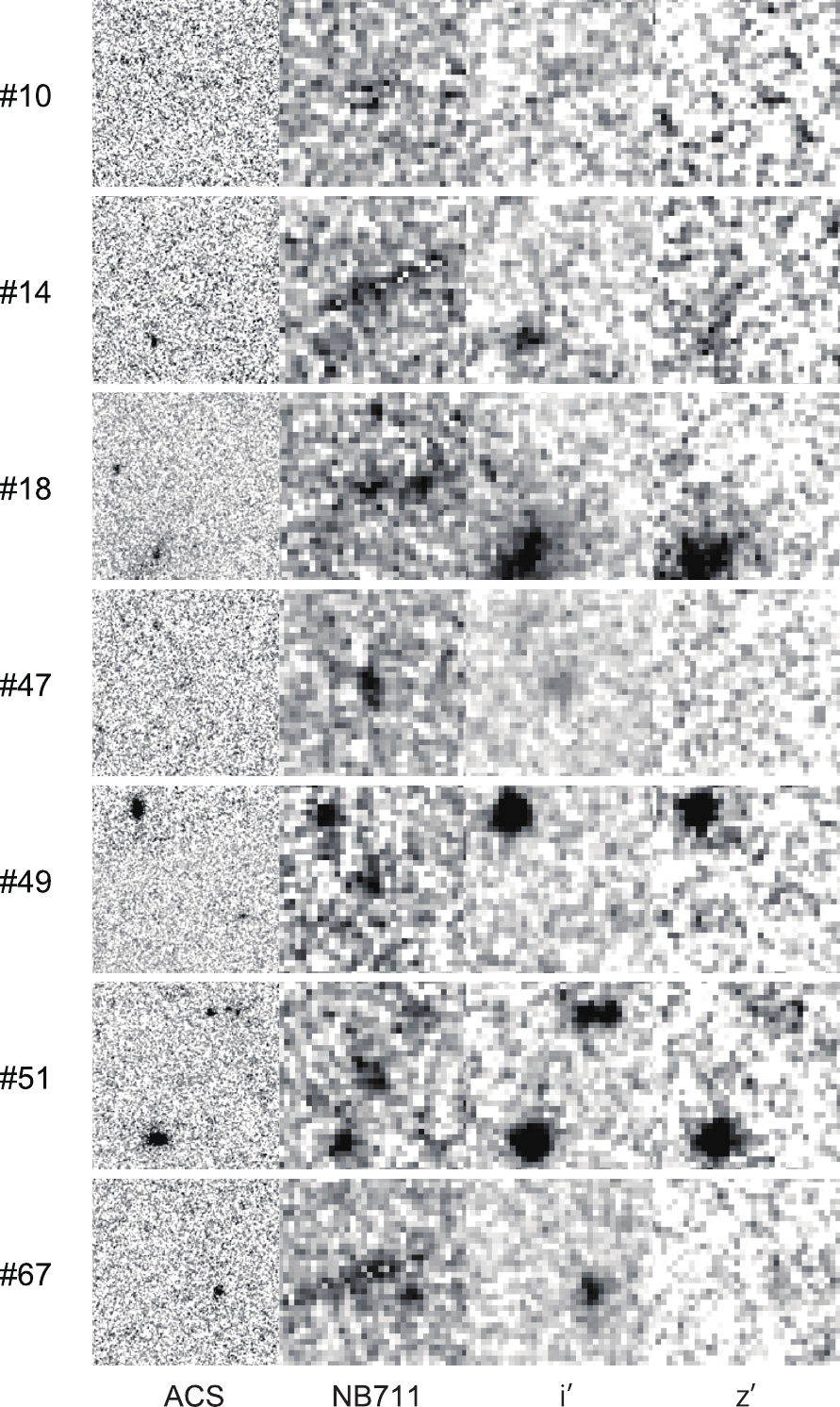}
  \caption{Same as Figure~\ref{fig:ThumbnailsDouble} but for 7 LAEs
  without ACS sources. \label{fig:ThumbnailsNot}}
 \end{figure}
 \begin{turnpage}
  \begin{deluxetable*}{lcccccccrrrc}
  \tabletypesize{\scriptsize}
  \tablecaption{\textit{ACS} F814W-band Properties for the 61 LAEs at
  $z = 4.86$ with ACS Data \label{tab:z4p9LAE}}
  \tablewidth{\linewidth}
  \tablehead{
  \colhead{ID\tablenotemark{a}} &
  \colhead{$I_{814}$\tablenotemark{b}} &
  \colhead{$a_\mathrm{HL}$\tablenotemark{c}} &
  \colhead{$R_\mathrm{HL}$\tablenotemark{d}} &
  \colhead{$\epsilon (I_{814})$\tablenotemark{e}} &
  \colhead{$NB711$\tablenotemark{f}} &
  \colhead{$a_\mathrm{HL} (\mathrm{NB711})$\tablenotemark{g}} &
  \colhead{$i^\prime$\tablenotemark{f}} &
  \colhead{$z^\prime$\tablenotemark{f}} &
  \colhead{$L(\mathrm{Ly\alpha})$\tablenotemark{h}} &
  \colhead{$\mathrm{EW_0}$\tablenotemark{i}} &
  \colhead{$z_\mathrm{spec}$\tablenotemark{j}}\\
  \colhead{} &
  \colhead{(mag)} &
  \colhead{(arcsec)} &
  \colhead{(arcsec)} &
  \colhead{} &
  \colhead{(mag)} &
  \colhead{(arcsec)} &
  \colhead{(mag)} &
  \colhead{(mag)} &
  \colhead{($10^{42}~\mathrm{erg~s^{-1}}$)} &
  \colhead{({\AA})} &
  \colhead{}
  }
  \startdata
 \multicolumn{12}{c}{8 ACS-detected LAEs with Double Components}\vspace{1mm}\\
 \cline{1-12}\vspace{-1mm}\\
  11 & $25.52 \pm 0.06$ & $0.49 \pm 0.02$ & $0.32 \pm 0.02$ & $0.76 \pm 0.01$ & $24.55 \pm 0.18$ & $0.51$ & $25.73 \pm 0.42$ & $24.98 \pm 0.24$ & $ 5.6 \pm 1.3$ & $18 \pm  6$ & $-99.0$\\
  12 & $24.11 \pm 0.03$ & $0.44 \pm 0.02$ & $0.34 \pm 0.01$ & $0.47 \pm 0.02$ & $23.64 \pm 0.08$ & $1.02$ & $24.16 \pm 0.07$ & $24.31 \pm 0.13$ & $ 9.5 \pm 1.2$ & $16 \pm  3$ & 4.850\\
  19 & $25.06 \pm 0.04$ & $0.55 \pm 0.02$ & $0.37 \pm 0.02$ & $0.79 \pm 0.01$ & $24.18 \pm 0.13$ & $0.47$ & $24.90 \pm 0.10$ & $24.82 \pm 0.17$ & $ 6.4 \pm 1.1$ & $18 \pm  4$ & \nodata\\
  21 & $25.05 \pm 0.04$ & $0.56 \pm 0.04$ & $0.24 \pm 0.01$ & $0.77 \pm 0.01$ & $23.96 \pm 0.12$ & $1.05$ & $25.26 \pm 0.12$ & $24.83 \pm 0.18$ & $10.1 \pm 1.3$ & $28 \pm  6$ & \nodata\\
  30 & $23.87 \pm 0.01$ & $0.31 \pm 0.00$ & $0.25 \pm 0.01$ & $0.54 \pm 0.01$ & $23.50 \pm 0.07$ & $0.56$ & $24.17 \pm 0.05$ & $24.13 \pm 0.09$ & $11.7 \pm 1.1$ & $17 \pm  2$ & \nodata\\
  37 & $25.21 \pm 0.06$ & $0.34 \pm 0.02$ & $0.23 \pm 0.01$ & $0.64 \pm 0.02$ & $24.08 \pm 0.13$ & $0.76$ & $24.59 \pm 0.07$ & $24.31 \pm 0.11$ & $ 5.8 \pm 1.2$ & $10 \pm  2$ & \nodata\\
  59 & $25.47 \pm 0.07$ & $0.59 \pm 0.05$ & $0.32 \pm 0.03$ & $0.81 \pm 0.02$ & $24.52 \pm 0.17$ & $0.54$ & $25.52 \pm 0.14$ & $26.03 \pm 0.47$ & $ 5.4 \pm 1.1$ & $45 \pm 26$ & \nodata\\
  68 & $25.02 \pm 0.02$ & $0.25 \pm 0.01$ & $0.13 \pm 0.00$ & $0.65 \pm 0.01$ & $24.06 \pm 0.12$ & $0.86$ & $24.65 \pm 0.07$ & $25.12 \pm 0.22$ & $ 6.5 \pm 1.1$ & $24 \pm  7$ & 4.798\vspace{1mm}\\
 \cline{1-12}\vspace{-2mm}\\
  \multicolumn{12}{c}{46 ACS-detected LAEs with Single Component}\vspace{1mm}\\
 \cline{1-12}\vspace{-1mm}\\
  1  & $26.20 \pm 0.10$ & $0.17 \pm 0.01$ & $0.13 \pm 0.01$ & $0.49 \pm 0.04$ & $24.86 \pm 0.26$ & $0.40$ & $26.54 \pm 0.32$ & $25.95 \pm 0.46$ & $ 4.8 \pm 1.2$ & $ 38 \pm 22$  & $-99.0$\\
  3  & $26.79 \pm 0.11$ & $0.11 \pm 0.01$ & $0.10 \pm 0.01$ & $0.13 \pm 0.11$ & $24.46 \pm 0.20$ & $0.37$ & $26.38 \pm 0.31$ & $> 26.64$        & $ 7.1 \pm 1.3$ & $>105$        & \nodata\\
  5  & $26.27 \pm 0.07$ & $0.10 \pm 0.01$ & $0.10 \pm 0.01$ & $0.13 \pm 0.07$ & $23.43 \pm 0.07$ & $0.53$ & $25.63 \pm 0.15$ & $25.61 \pm 0.29$ & $19.0 \pm 1.1$ & $109 \pm  34$ & 4.839\\
  9  & $27.30 \pm 0.10$ & $0.08 \pm 0.01$ & $0.06 \pm 0.00$ & $0.37 \pm 0.09$ & $24.22 \pm 0.13$ & $0.58$ & $26.02 \pm 0.21$ & $> 26.64$        & $ 8.8 \pm 1.1$ & $> 130$       & \nodata\\
  13 & $26.97 \pm 0.07$ & $0.09 \pm 0.00$ & $0.08 \pm 0.00$ & $0.24 \pm 0.08$ & $24.71 \pm 0.20$ & $0.44$ & $25.97 \pm 0.19$ & $25.67 \pm 0.31$ & $ 5.0 \pm 1.1$ & $ 30 \pm  12$ & \nodata\\
  15 & $25.73 \pm 0.06$ & $0.12 \pm 0.01$ & $0.12 \pm 0.01$ & $0.06 \pm 0.06$ & $24.07 \pm 0.12$ & $1.02$ & $25.18 \pm 0.11$ & $25.43 \pm 0.27$ & $ 8.7 \pm 1.2$ & $ 42 \pm  13$ & \nodata\\
  16 & $27.26 \pm 0.16$ & $0.09 \pm 0.01$ & $0.08 \pm 0.02$ & $0.13 \pm 0.14$ & $24.46 \pm 0.17$ & $0.60$ & $25.43 \pm 0.13$ & $25.87 \pm 0.39$ & $ 5.4 \pm 1.1$ & $ 39 \pm  19$ & \nodata\\
  17 & $27.09 \pm 0.16$ & $0.12 \pm 0.02$ & $0.11 \pm 0.01$ & $0.13 \pm 0.16$ & $24.74 \pm 0.22$ & $0.59$ & $26.76 \pm 0.40$ & $> 26.64$        & $ 5.6 \pm 1.2$ & $>  83$       & \nodata\\
  22 & $27.00 \pm 0.11$ & $0.09 \pm 0.01$ & $0.08 \pm 0.01$ & $0.21 \pm 0.08$ & $24.51 \pm 0.17$ & $0.90$ & $26.36 \pm 0.24$ & $> 26.64$        & $ 6.8 \pm 2.8$ & $> 101$       & \nodata\\
  23 & $25.33 \pm 0.04$ & $0.23 \pm 0.00$ & $0.16 \pm 0.01$ & $0.56 \pm 0.02$ & $24.41 \pm 0.15$ & $0.42$ & $24.94 \pm 0.08$ & $24.73 \pm 0.15$ & $ 4.5 \pm 1.0$ & $ 11 \pm   3$ & \nodata\\
  24 & $26.54 \pm 0.10$ & $0.12 \pm 0.01$ & $0.11 \pm 0.01$ & $0.24 \pm 0.06$ & $24.49 \pm 0.18$ & $0.62$ & $26.12 \pm 0.18$ & $> 26.64$        & $ 6.7 \pm 1.2$ & $>  98$       & 4.845\\
  25 & $26.78 \pm 0.11$ & $0.10 \pm 0.01$ & $0.08 \pm 0.01$ & $0.27 \pm 0.08$ & $24.38 \pm 0.17$ & $0.80$ & $25.04 \pm 0.10$ & $24.49 \pm 0.13$ & $ 5.1 \pm 1.3$ & $ 10 \pm   3$ & \nodata\\
  26 & $26.18 \pm 0.07$ & $0.11 \pm 0.01$ & $0.10 \pm 0.00$ & $0.20 \pm 0.05$ & $24.46 \pm 0.17$ & $0.42$ & $26.12 \pm 0.21$ & $> 26.64$        & $ 6.8 \pm 1.1$ & $> 101$       & \nodata\\
  27 & $26.62 \pm 0.07$ & $0.15 \pm 0.01$ & $0.11 \pm 0.00$ & $0.49 \pm 0.05$ & $24.38 \pm 0.15$ & $0.78$ & $25.88 \pm 0.18$ & $25.73 \pm 0.35$ & $ 7.3 \pm 1.1$ & $ 47 \pm  19$ & \nodata\\
  29 & $26.70 \pm 0.04$ & $0.08 \pm 0.00$ & $0.07 \pm 0.00$ & $0.20 \pm 0.05$ & $23.87 \pm 0.10$ & $0.65$ & $26.82 \pm 0.39$ & $> 26.64$        & $13.4 \pm 1.1$ & $> 199$       & \nodata\\
  31 & $25.98 \pm 0.07$ & $0.14 \pm 0.01$ & $0.12 \pm 0.01$ & $0.26 \pm 0.05$ & $24.58 \pm 0.18$ & $0.65$ & $26.13 \pm 0.22$ & $25.48 \pm 0.31$ & $ 6.1 \pm 1.1$ & $ 31 \pm  12$ & \nodata\\
  33 & $26.17 \pm 0.08$ & $0.13 \pm 0.01$ & $0.11 \pm 0.01$ & $0.33 \pm 0.05$ & $24.64 \pm 0.21$ & $0.51$ & $26.13 \pm 0.26$ & $> 26.64$        & $ 5.6 \pm 1.3$ & $>  83$       & \nodata\\
  34 & $24.61 \pm 0.03$ & $0.15 \pm 0.00$ & $0.13 \pm 0.00$ & $0.27 \pm 0.02$ & $23.56 \pm 0.08$ & $0.70$ & $24.78 \pm 0.08$ & $25.14 \pm 0.21$ & $14.0 \pm 1.2$ & $ 52 \pm  12$ & \nodata\\
  38 & $25.82 \pm 0.06$ & $0.13 \pm 0.01$ & $0.11 \pm 0.01$ & $0.29 \pm 0.05$ & $24.04 \pm 0.11$ & $0.51$ & $26.37 \pm 0.22$ & $> 26.64$        & $11.0 \pm 1.1$ & $> 163$       & 4.873\\
  39 & $25.94 \pm 0.12$ & $0.20 \pm 0.02$ & $0.16 \pm 0.03$ & $0.59 \pm 0.04$ & $24.21 \pm 0.19$ & $0.87$ & $26.05 \pm 0.19$ & $> 26.64$        & $ 8.8 \pm 1.6$ & $> 131$       & $-99.0$\\
  40 & $27.45 \pm 0.15$ & $0.09 \pm 0.01$ & $0.09 \pm 0.01$ & $0.04 \pm 0.16$ & $24.76 \pm 0.20$ & $0.51$ & $26.05 \pm 0.17$ & $26.26 \pm 0.55$ & $ 4.8 \pm 1.0$ & $ 50 \pm  34$ & 4.818\\
  41 & $24.97 \pm 0.04$ & $0.09 \pm 0.00$ & $0.09 \pm 0.00$ & $0.04 \pm 0.04$ & $23.39 \pm 0.07$ & $0.47$ & $24.88 \pm 0.08$ & $25.62 \pm 0.36$ & $17.7 \pm 1.2$ & $103 \pm  41$ & 4.830\\
  42 & $26.05 \pm 0.07$ & $0.18 \pm 0.01$ & $0.13 \pm 0.01$ & $0.51 \pm 0.04$ & $23.82 \pm 0.09$ & $0.75$ & $25.37 \pm 0.11$ & $25.58 \pm 0.30$ & $12.1 \pm 1.0$ & $ 67 \pm  22$ & \nodata\\
  43 & $26.41 \pm 0.11$ & $0.13 \pm 0.01$ & $0.11 \pm 0.01$ & $0.43 \pm 0.07$ & $24.49 \pm 0.16$ & $0.79$ & $25.83 \pm 0.16$ & $25.59 \pm 0.31$ & $ 6.1 \pm 1.1$ & $ 34 \pm  13$ & \nodata\\
  44 & $25.63 \pm 0.04$ & $0.12 \pm 0.00$ & $0.11 \pm 0.00$ & $0.24 \pm 0.04$ & $23.91 \pm 0.10$ & $9999.0$ & $24.93 \pm 0.10$ & $25.16 \pm 0.21$ & $ 9.4 \pm 1.1$ & $ 36 \pm   9$ & \nodata\\
  45 & $27.07 \pm 0.17$ & $0.17 \pm 0.04$ & $0.12 \pm 0.02$ & $0.56 \pm 0.08$ & $24.62 \pm 0.16$ & $0.49$ & $26.01 \pm 0.16$ & $26.54 \pm 0.83$ & $ 5.5 \pm 1.0$ & $ 74 \pm  69$ & 4.865\\
  46 & $25.76 \pm 0.05$ & $0.10 \pm 0.01$ & $0.09 \pm 0.00$ & $0.17 \pm 0.05$ & $24.65 \pm 0.17$ & $0.61$ & $25.48 \pm 0.11$ & $25.27 \pm 0.23$ & $ 4.5 \pm 1.0$ & $ 19 \pm   6$ & 4.865\\
  48 & $26.65 \pm 0.09$ & $0.07 \pm 0.00$ & $0.07 \pm 0.00$ & $0.07 \pm 0.07$ & $23.95 \pm 0.11$ & $0.52$ & $26.09 \pm 0.19$ & $> 26.64$        & $11.7 \pm 1.1$ & $> 173$       & \nodata\\
  50 & $25.99 \pm 0.06$ & $0.09 \pm 0.00$ & $0.09 \pm 0.00$ & $0.02 \pm 0.06$ & $23.69 \pm 0.09$ & $0.74$ & $26.22 \pm 0.25$ & $26.34 \pm 0.65$ & $15.4 \pm 1.2$ & $172 \pm 143$ & \nodata\\
  54 & $25.35 \pm 0.03$ & $0.08 \pm 0.00$ & $0.07 \pm 0.00$ & $0.07 \pm 0.03$ & $23.45 \pm 0.07$ & $0.64$ & $25.42 \pm 0.13$ & $25.44 \pm 0.29$ & $18.3 \pm 1.1$ & $ 90 \pm  28$ & \nodata\\
  55 & $25.09 \pm 0.05$ & $0.17 \pm 0.01$ & $0.13 \pm 0.01$ & $0.40 \pm 0.03$ & $24.50 \pm 0.15$ & $0.75$ & $25.21 \pm 0.10$ & $25.22 \pm 0.23$ & $ 4.8 \pm 1.0$ & $ 19 \pm   6$ & 4.830\\
  56 & $25.50 \pm 0.05$ & $0.19 \pm 0.01$ & $0.15 \pm 0.01$ & $0.54 \pm 0.02$ & $24.45 \pm 0.16$ & $0.71$ & $25.42 \pm 0.11$ & $26.08 \pm 0.50$ & $ 5.6 \pm 1.1$ & $ 49 \pm  31$ & $-99.0$\\
  57 & $26.91 \pm 0.10$ & $0.12 \pm 0.01$ & $0.09 \pm 0.01$ & $0.42 \pm 0.07$ & $24.67 \pm 0.21$ & $0.33$ & $26.53 \pm 0.33$ & $25.54 \pm 0.30$ & $ 5.9 \pm 1.2$ & $ 32 \pm  12$ & \nodata\\
  58 & $25.79 \pm 0.06$ & $0.15 \pm 0.01$ & $0.13 \pm 0.01$ & $0.29 \pm 0.04$ & $23.90 \pm 0.12$ & $0.85$ & $25.72 \pm 0.15$ & $25.50 \pm 0.31$ & $11.8 \pm 1.3$ & $ 61 \pm  21$ & 4.840\\
  60 & $25.89 \pm 0.06$ & $0.14 \pm 0.01$ & $0.13 \pm 0.01$ & $0.13 \pm 0.06$ & $24.14 \pm 0.14$ & $0.70$ & $25.43 \pm 0.13$ & $25.28 \pm 0.24$ & $ 8.5 \pm 1.2$ & $ 36 \pm  10$ & \nodata\\
  61 & $26.84 \pm 0.10$ & $0.14 \pm 0.01$ & $0.12 \pm 0.01$ & $0.33 \pm 0.07$ & $24.80 \pm 0.22$ & $0.31$ & $26.09 \pm 0.22$ & $25.71 \pm 0.34$ & $ 4.6 \pm 1.1$ & $ 29 \pm  13$ & \nodata\\
  63 & $25.79 \pm 0.04$ & $0.10 \pm 0.00$ & $0.08 \pm 0.00$ & $0.31 \pm 0.03$ & $23.68 \pm 0.09$ & $0.55$ & $24.31 \pm 0.06$ & $23.91 \pm 0.08$ & $ 9.2 \pm 1.2$ & $ 11 \pm   2$ & \nodata\\
  64 & $25.99 \pm 0.09$ & $0.15 \pm 0.01$ & $0.12 \pm 0.01$ & $0.40 \pm 0.05$ & $24.27 \pm 0.14$ & $0.75$ & $25.09 \pm 0.10$ & $25.30 \pm 0.28$ & $ 6.1 \pm 1.1$ & $ 26 \pm   9$ & \nodata\\
  65 & $25.24 \pm 0.06$ & $0.14 \pm 0.01$ & $0.12 \pm 0.01$ & $0.21 \pm 0.04$ & $24.41 \pm 0.16$ & $0.61$ & $25.06 \pm 0.10$ & $24.81 \pm 0.16$ & $ 5.0 \pm 1.1$ & $ 14 \pm   4$ & \nodata
  \enddata
  \end{deluxetable*}
 \end{turnpage}
 \setcounter{table}{2}
 \begin{turnpage}
  \begin{deluxetable*}{lcccccccrrrc}
  \tabletypesize{\scriptsize}
  \tablecaption{(Continued.)}
  \tablewidth{\linewidth}
  \tablehead{
  \colhead{ID\tablenotemark{a}} &
  \colhead{$I_{814}$\tablenotemark{b}} &
  \colhead{$a_\mathrm{HL}$\tablenotemark{c}} &
  \colhead{$R_\mathrm{HL}$\tablenotemark{d}} &
  \colhead{$\epsilon (I_{814})$\tablenotemark{e}} &
  \colhead{$NB711$\tablenotemark{f}} &
  \colhead{$a_\mathrm{HL} (\mathrm{NB711})$\tablenotemark{g}} &
  \colhead{$i^\prime$\tablenotemark{f}} &
  \colhead{$z^\prime$\tablenotemark{f}} &
  \colhead{$L(\mathrm{Ly\alpha})$\tablenotemark{h}} &
  \colhead{$\mathrm{EW_0}$\tablenotemark{i}} &
  \colhead{$z_\mathrm{spec}$\tablenotemark{j}}\\
  \colhead{} &
  \colhead{(mag)} &
  \colhead{(arcsec)} &
  \colhead{(arcsec)} &
  \colhead{} &
  \colhead{(mag)} &
  \colhead{(arcsec)} &
  \colhead{(mag)} &
  \colhead{(mag)} &
  \colhead{($10^{42}~\mathrm{erg~s^{-1}}$)} &
  \colhead{({\AA})} &
  \colhead{}
  }
  \startdata
  66 & $26.84 \pm 0.10$ & $0.13 \pm 0.01$ & $0.10 \pm 0.01$ & $0.34 \pm 0.08$ & $24.88 \pm 0.26$ & $0.44$ & $26.84 \pm 0.42$ & $> 26.64$        & $ 4.9 \pm 1.3$ & $>  72$     & \nodata\\
  69 & $27.57 \pm 0.12$ & $0.07 \pm 0.01$ & $0.06 \pm 0.01$ & $0.15 \pm 0.15$ & $24.49 \pm 0.15$ & $0.58$ & $25.20 \pm 0.10$ & $25.79 \pm 0.39$ & $ 4.5 \pm 1.0$ & $31 \pm 15$ & 4.854\\
  70 & $25.09 \pm 0.03$ & $0.13 \pm 0.00$ & $0.10 \pm 0.00$ & $0.34 \pm 0.03$ & $23.50 \pm 0.07$ & $0.47$ & $24.67 \pm 0.07$ & $24.88 \pm 0.17$ & $14.6 \pm 1.1$ & $43 \pm  8$ & \nodata\\
  71 & $25.27 \pm 0.05$ & $0.15 \pm 0.01$ & $0.13 \pm 0.00$ & $0.22 \pm 0.04$ & $24.32 \pm 0.14$ & $0.52$ & $24.87 \pm 0.08$ & $24.66 \pm 0.14$ & $ 4.9 \pm 1.1$ & $12 \pm  3$ & \nodata\\
  72 & $25.51 \pm 0.08$ & $0.30 \pm 0.01$ & $0.22 \pm 0.01$ & $0.50 \pm 0.03$ & $24.35 \pm 0.15$ & $0.54$ & $25.59 \pm 0.15$ & $> 26.64$        & $ 7.0 \pm 1.1$ & $> 104$     & \nodata\\
  77 & $26.56 \pm 0.11$ & $0.09 \pm 0.01$ & $0.08 \pm 0.01$ & $0.21 \pm 0.07$ & $24.70 \pm 0.22$ & $0.37$ & $26.19 \pm 0.25$ & $> 26.64$        & $ 5.4 \pm 1.2$ & $>  81$     & \nodata\\
  78 & $25.60 \pm 0.05$ & $0.15 \pm 0.01$ & $0.13 \pm 0.01$ & $0.24 \pm 0.04$ & $23.25 \pm 0.06$ & $0.55$ & $25.05 \pm 0.10$ & $24.83 \pm 0.16$ & $21.2 \pm 1.2$ & $59 \pm 10$ & \nodata\vspace{1mm}\\
  \cline{1-12}\vspace{-1mm}\\
  \multicolumn{12}{c}{7 ACS-undetected LAEs}\vspace{1mm}\\
  \cline{1-12}\vspace{-1mm}\\
  10 & \nodata & \nodata & \nodata & \nodata & $24.80 \pm 0.22$ & $0.41$ & $27.04 \pm 0.61$ & $> 26.64$        & $5.5 \pm 1.2$ & $>  82$     & $-99.0$\\
  14 & \nodata & \nodata & \nodata & \nodata & $24.22 \pm 0.13$ & $1.06$ & $26.99 \pm 0.58$ & $> 26.64$        & $9.6 \pm 1.1$ & $> 142$     & \nodata\\
  18 & \nodata & \nodata & \nodata & \nodata & $24.10 \pm 0.12$ & $1.64$ & $25.23 \pm 0.11$ & $25.25 \pm 0.22$ & $8.2 \pm 1.1$ & $34 \pm  9$ & \nodata\\
  47 & \nodata & \nodata & \nodata & \nodata & $24.70 \pm 0.19$ & $0.60$ & $26.60 \pm 0.27$ & $26.49 \pm 0.64$ & $5.7 \pm 1.0$ & $74 \pm 60$ & 4.840\\
  49 & \nodata & \nodata & \nodata & \nodata & $24.93 \pm 0.33$ & $0.35$ & $26.72 \pm 0.38$ & $25.84 \pm 0.44$ & $4.6 \pm 1.6$ & $33 \pm 20$ & \nodata\\
  51 & \nodata & \nodata & \nodata & \nodata & $24.64 \pm 0.21$ & $0.46$ & $25.56 \pm 0.16$ & $25.67 \pm 0.36$ & $4.5 \pm 1.3$ & $27 \pm 13$ & \nodata\\
  67 & \nodata & \nodata & \nodata & \nodata & $24.16 \pm 0.13$ & $0.99$ & $26.16 \pm 0.20$ & $> 26.64$        & $9.5 \pm 1.1$ & $> 141$     & \nodata
  \enddata

  \tablecomments{(a) The LAE ID given in Shioya et al.~(2009).  (b)
  SExtractor's MAG\_AUTO magnitude and its $1\sigma$ error.  (c)
  Half-light major radius and its $1\sigma$ error measured on ACS
  F814W-band images.  (d) Half-light radius and its $1\sigma$ error
  measured on ACS F814W-band images.  (e) Ellipticity and its
  $1\sigma$ error measured on ACS F814W-band images.  (f)
  $3^{\prime\prime}$ diameter aperture magnitude and its $1\sigma$
  error.  (g) Half-light major radius measured on NB711--band images.
  The entry of 9999.0 for the LAE \#44 means that its size estimation
  is impossible because of the presence of a close bright contaminant.
  (h) Ly$\alpha$ line luminosity and its $1\sigma$ error.  (i)
  Rest-frame Ly$\alpha$ EW and its $1\sigma$ error.  Note that these
  values are different from $\mathrm{EW_0}$ listed in Table~1 in S09
  by a factor of 0.83 because of an error (see Erratum of S09).  (j)
  Spectroscopic redshift.  The entry of $-99.0$ means that redshift is
  not determined whereas follow-up spectroscopy is performed.}

  \end{deluxetable*}
 \end{turnpage}
 We show the thumbnails of the 61 LAEs in the ACS F814W-band images
 together with their Subaru NB711--, $i^\prime$--, and
 $z^\prime$--band images in Figure~\ref{fig:ThumbnailsDouble} (8
 ACS-detected LAEs with double-component),
 Figure~\ref{fig:ThumbnailsSingle} (46 ACS-detected LAEs with
 single-component), and Figure~\ref{fig:ThumbnailsNot} (7
 ACS-undetected LAEs).  In these figures, the detected ACS sources
 identified as LAE counterparts are indicated by red ellipses on the
 NB711--, $i^\prime$--, and $z^\prime$--band images.  For the
 double-component LAEs shown in Figure~\ref{fig:ThumbnailsDouble}, the
 individual ACS sources detected are also overlayed by yellow ellipses
 on the NB711--, $i^\prime$--, and $z^\prime$--band images.

 The total magnitude ($I_{814}$), circularized half-light radius
 ($R_\mathrm{HL}$), half-light major radius ($a_\mathrm{HL}$), and
 ellipticity ($\epsilon$) are measured for each detected source with
 SExtractor on the original ACS F814W-band image (i.e., not on the
 smoothed image).  We cannot use the profile fitting which is usually
 used to estimate the radius and ellipticity because it is not obvious
 whether or not the profile fitting can estimate intrinsic radius and
 ellipticity well for very faint sources like our sources which is
 fainter than previous studies.  The ellipticity is defined as
 $\epsilon = 1 - b / a$, where $a$ and $b$ are the major and minor
 radii, respectively.  We adopt SExtractor's MAG\_AUTO, MAGERR\_AUTO,
 and FLUX\_RADIUS with PHOT\_FLUXFRAC of 0.5 as $I_{814}$, error of
 $I_{814}$ and $R_\mathrm{HL}$, respectively.  In order to obtain
 half-light major radius $a_{\rm HL}$, we modified the code for
 growth-curve measurement (growth.c) in SExtractor so that the
 half-light radius is measured with elliptical apertures which have
 the same ellipticity and position angle derived from the second-order
 moments by the SExtractor rather than circular apertures.  For the
 double-component LAEs as single sources, these properties are
 evaluated using both of SExtractor and IDL.  The Errors of
 $a_\mathrm{HL}$ and $R_\mathrm{HL}$ are based on the magnitude error.
 The errors of ellipticity is based on local background noise
 fluctuation.

 These photometric properties of the ACS data are listed in
 Table~\ref{tab:z4p9LAE}.  Note that the 3$\sigma$ limiting magnitude
 of the F814W-band images is 27.4~mag in a $1^{\prime\prime}$ diameter
 aperture.  All magnitudes are corrected for the Galactic extinction
 of $A_\mathrm{F814W} = 0.035$ (Capak et al. 2007).  In
 Table~\ref{tab:z4p9LAE}, we also list the photometric properties of
 the LAE candidates from S09.  The 3$\sigma$ limiting magnitudes
 within a $3^{\prime\prime}$ diameter aperture in the NB711--,
 $i^\prime$--, and $z^\prime$--band images are 25.17, 26.49, and
 25.45, respectively.

 \section{MORPHOLOGICAL PROPERTIES}

 The ACS counterparts of the LAEs look differently from object to
 object as shown in Figures~\ref{fig:ThumbnailsDouble} and
 \ref{fig:ThumbnailsSingle}.  Here we examine first which emission the
 ACS F814W-band image probes, Ly$\alpha$ line or UV stellar continuum.
 Then we present the morphological properties of the 54 ACS-detected
 LAEs measured on the ACS F814W-band image, that is, half-light radius
 $R_\mathrm{HL}$, half-light major radius $a_\mathrm{HL}$, and
 ellipticity $\epsilon$.

  \subsection{What Do ACS F814W-band Images Probe?}\label{subsec:ACS}

  As presented in Figure~\ref{fig:filters}, the transmission curve of
  the F814W-band filter covers both Ly$\alpha$ line emission and
  rest-frame UV continuum emission at wavelengths of $\sim
  1200$--1640~{\AA} from a source at $z = 4.86$.  Which emission do
  ACS F814W-band images mainly probe?

  \begin{figure*}
   \plotone{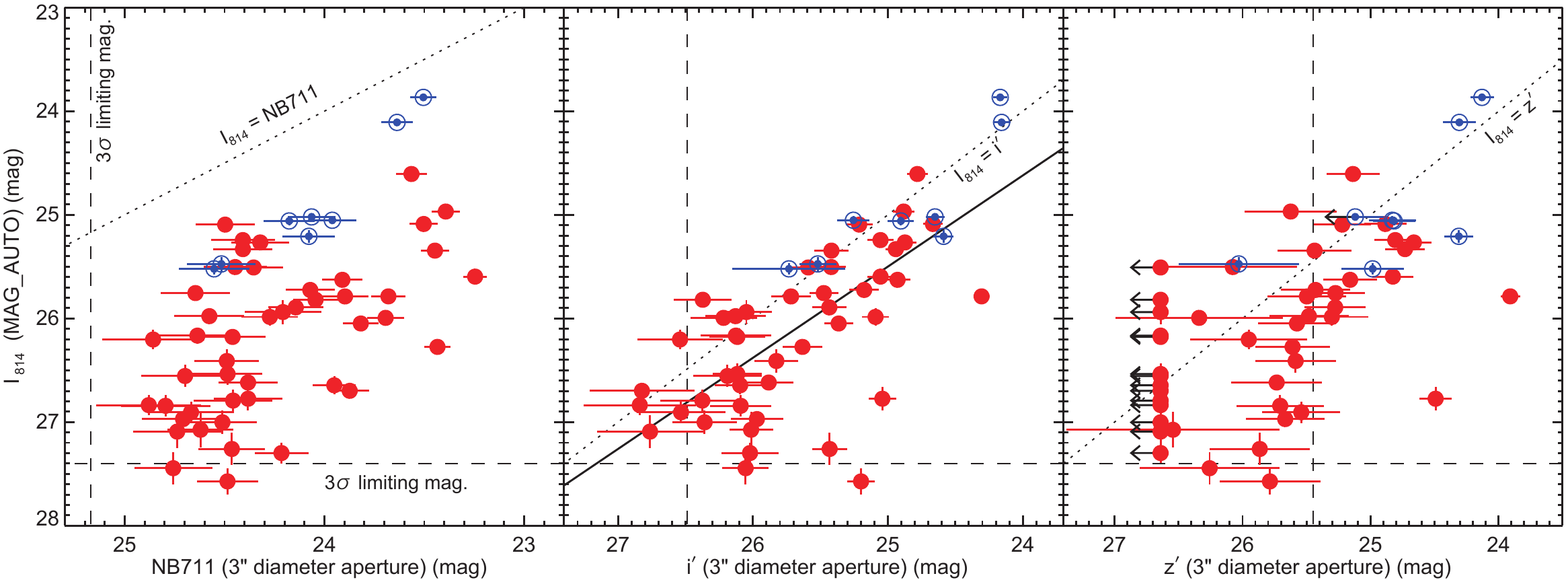}

   \caption{Distribution of the 62 ACS-detected LAEs in the
   $I_{814}$--$NB711$ (left), $I_{814}$--$i^\prime$ (middle), and
   $I_{814}$--$z^\prime$ planes (right).  For the double-component
   LAEs represented by blue double circles, sum of $I_{814}$ of each
   component is adopted in this plot.  The dotted lines represent the
   equality of $I_{814}$ and $NB711$, $i^\prime$, or $z^\prime$.  The
   $3\sigma$ limiting magnitude of ACS in a $1^{\prime\prime}$
   diameter aperture, 27.4~mag, is shown by the horizontal dashed
   line, while those of NB711--, $i^\prime$--, and $z^\prime$--band
   images in a $3^{\prime\prime}$ diameter aperture, 25.17, 26.49, and
   25.45~mag, are shown by the vertical dashed lines, respectively.
   In the middle panel, the solid line represents the best-fit linear
   relation between $I_{814}$ and $i^\prime$: $I_{814} = 0.88(i^\prime
   - 29.0) + 29.0$.  In the right panel, the LAEs fainter than the
   $1\sigma$ limiting magnitude of $z^\prime$ are located at $z^\prime
   = z^\prime(1\sigma) = 26.64$~mag with
   arrows. \label{fig:I814vsNB711}}

  \end{figure*}
  The detected emission in the F814W-band filter seems to be primarily
  from rest-frame UV continuum rather than Ly$\alpha$ line emission.
  This is clearly exhibited by the presence of a positive correlation
  between $I_{814}$ and $i^\prime$, which are close to $I_{814} \sim
  i^\prime$, as shown in the middle panel of
  Figure~\ref{fig:I814vsNB711}.  The linear correlation coefficient is
  estimated to be $r = 0.72$.  It is similar for $z^\prime$, which
  linear correlation coefficient is $r = 0.61$\footnote{In this
  calculation, the LAEs with $z^\prime \ge z^\prime(1\sigma) =
  26.64$~mag are excluded.}, while dispersion from $I_{814} =
  z^\prime$ relation is more significant.  On the other hand, the
  correlation between $I_{814}$ and $NB711$ appears to be poorer
  compared with the correlations between $I_{814}$ and $i^\prime$ or
  $z^\prime$ (its linear correlation coefficient is $r = 0.56$).  This
  result is consistent with the facts that most LAEs have
  observer-frame EW much smaller than $\Delta \lambda$ of the F814W
  (i.e., 2511~{\AA}) and that the wavelength of the NB711 band which
  is almost blue edge of the wavelength coverage of the F814W-band
  filter (see Figure~\ref{fig:filters}).

  Therefore, we can conclude that the ACS F814W-band images primarily
  probe rest-frame UV continuum emission from young massive stars in
  the LAEs at $z = 4.86$.

  \subsection{Size: Half-light Radius and Half-light Major
  Radius}\label{subsec:size}

  Then we analyze the sizes of our LAE sample in the ACS F814W-band
  images, that is, half-light radius $R_\mathrm{HL}$ and half-light
  major radius $a_\mathrm{HL}$.  We emphasize that these measured
  sizes should be considered as the extent of the young star-forming
  regions in the LAEs and they do not necessarily reflect the stellar
  mass distribution since the F814W-band images mainly prove their
  rest-frame UV continuum emissions at wavelengths of $\sim
  1200$--1640~{\AA} as presented in
  Section~\ref{subsec:ACS}\footnote{The Wide Fields Camera 3 (WFC3)
  F160W-band images ($\lambda_c = 15,369$~{\AA} and $\Delta \lambda =
  2683$~{\AA}) are also available only in a limited part of the COSMOS
  field (210~$\mathrm{arcmin^2} \approx 3\%$ of the COSMOS field),
  which are taken by the Cosmic Assembly Near-infrared Deep
  Extragalactic Legacy Survey (CANDELS; Grogin et al. 2011; Koekemoer
  et al. 2011).  Although the F160W-band images can prove our LAE
  samples at the slightly longer rest-frame wavelengths of $\sim
  2390$--2850~{\AA}, only two single-component LAEs of \#46 and \#55
  are covered in the CANDELS/COSMOS field; therefore, we do not show
  the morphological properties of our LAE sample in the F160W-band
  images in this paper.  We just comment that, while their sizes in
  the F160W-band images are larger than those in the F814W-band
  images, the differences of the sizes between these two-band images
  are consistent the differences of the PSF sizes and pixel scales.}.
  Note that the measured half-light radii of $\approx 4600$
  unsaturated stars with $I_{814} = 20$--22~mag and $\mathrm{FWHM} \le
  4~\mathrm{pix}$ ($= 0\farcs 12$) in the ACS F814W-band images,
  $R_\mathrm{PSF}$, are typically 0\farcs 07; we adopt this angular
  scale as the ``PSF size'' of the ACS F814W-band images\footnote{We
  also measure FWHMs of the same stars and obtain a typical FWHM of
  0\farcs 1, which is consistent with the average PSF FWHM reported by
  Koekemoer et al. (2007).  Note that the half-light radius
  $R_\mathrm{PSF}$ is smaller than the measured FWHMs of stars by a
  factor of 2 in the case that the PSF is completely described by
  Gaussian profile.  The actual PSF is different from a Gaussian
  profile and hence the ratio of $\mathrm{FWHM} / R_\mathrm{PSF}$ can
  be different from 2.  Since a confusion of $R_\mathrm{PSF}$ and FWHM
  for the term of ``PSF size'' is seen in a non-negligible number of
  literatures, we emphasized that a particular attention should be
  paid to which of $R_\mathrm{PSF}$ or FWHM the ``PSF size''
  indicates.} in our analysis.

  \begin{figure}
   \plotone{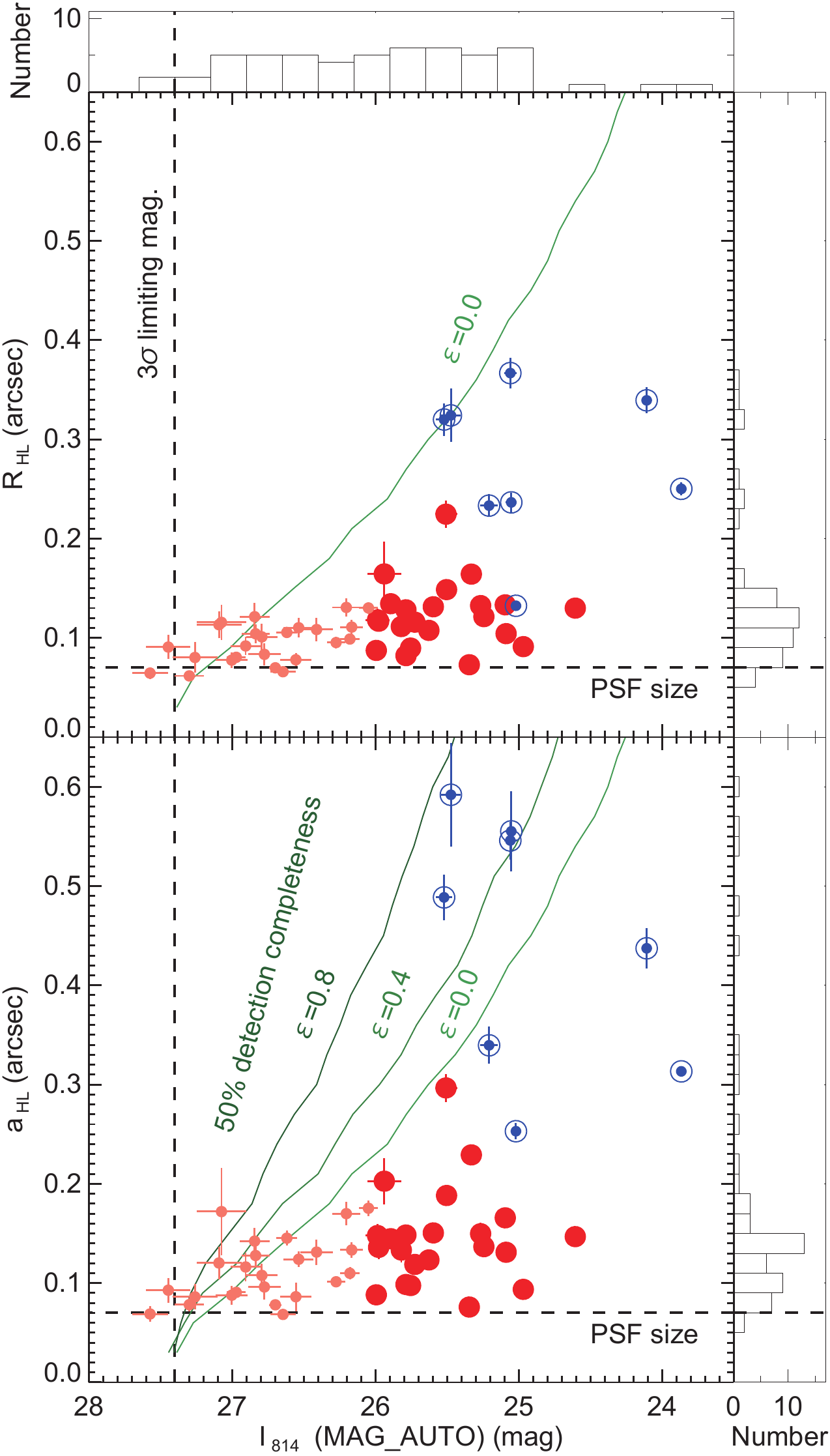}

   \caption{Distributions of the 54 ACS-detected LAEs in the
   $R_\mathrm{HL}$--$I_{814}$ (top) and $a_\mathrm{HL}$--$I_{814}$
   (bottom) planes.  In the main panels, the LAEs with single- and
   double-components in the ACS F814W-band images are represented by
   the red filled and blue double circles, respectively.  For the 8
   double-component LAEs, the size and magnitude of the double ACS
   components measured as a single object are plotted.  The ACS
   sources with $I_{814} \ge 26$~mag are shown by the small symbols.
   The $3\sigma$ limiting magnitude in a $1^{\prime\prime}$ diameter
   aperture and PSF half-light radius derived from stars, 27.4~mag and
   0\farcs 07, are also shown by the vertical and horizontal dashed
   lines, respectively.  Note that we show the same value of
   $R_\mathrm{PSF}$ both in the top and bottom panels since the PSF is
   found to have negligibly small ellipticity (i.e.,
   $\epsilon_\mathrm{PSF} < 0.03$).  The solid curves in the bottom
   panel indicate the 50\% detection completenesses for exponential
   disk objects with the input ellipticities of $\epsilon^\mathrm{in}
   = 0.0$, 0.4, and 0.8 estimated by a Monte Carlo simulation.  The
   same curve with $\epsilon^\mathrm{in} = 0.0$ is also depicted by
   the solid curve in the top panel.  The distributions shown in the
   main panels are projected onto the two side panels where histograms
   of $I_{814}$, $R_\mathrm{HL}$, and $a_\mathrm{HL}$ are
   displayed. \label{fig:size-I814}}

  \end{figure}
  Figure~\ref{fig:size-I814} shows the distributions of the 54
  ACS-detected LAEs in the $R_\mathrm{HL}$--$I_{814}$ and
  $a_\mathrm{HL}$--$I_{814}$ planes.  It is found that the ACS
  magnitudes $I_{814}$ of the LAEs are widely distributed in
  23.87--27.57~mag with the mean value of $\langle I_{814} \rangle =
  25.99 \pm 0.11$~mag.  Both sizes of $R_\mathrm{HL}$ and
  $a_\mathrm{HL}$ of the LAEs are also found to be widely distributed
  in $R_\mathrm{HL} = 0\farcs 06$--0\farcs 37 and $a_\mathrm{HL} =
  0\farcs 07$--0\farcs 59 with mean values of $\langle R_\mathrm{HL}
  \rangle = 0\farcs 133 \pm 0\farcs 010$ and $\langle a_\mathrm{HL}
  \rangle = 0\farcs 175 \pm 0\farcs 017$.  Their distributions are
  similar with each other, having a concentration at small sizes and
  an elongated tail toward large sizes.  As shown in
  Figure~\ref{fig:size-I814}, both distributions of $R_\mathrm{HL}$
  and $a_\mathrm{HL}$ are not concentrated around the means but show a
  clear separation between the single- and double-component LAEs; the
  latters have larger sizes than the formers typically.  Moreover, all
  single-component LAEs are found to have sizes of $\lesssim 0\farcs
  3$; if the double-component LAEs are excluded, the mean half-light
  major radius becomes 0\farcs 13.

  Most of the ACS sources have $a_\mathrm{HL} > R_\mathrm{HL}$,
  implying that they have non-zero ellipticities.  Since
  $a_\mathrm{HL}$ is generally considered to be a more appropriate
  measure of size than $R_\mathrm{HL}$ for such sources having
  non-zero ellipticities, we adopt $a_\mathrm{HL}$ as the fiducial
  size of the individual ACS source in the following, rather than
  $R_\mathrm{HL}$.

  In Figure~\ref{fig:size-I814}, in order to see the effect of
  limiting surface brightness in our ACS data, we also plot the
  50\% detection completeness limits for faint extended sources in the
  ACS F814W-band images estimated via performing Monte Carlo
  simulations; the details of our Monte Carlo simulations are
  described in Appendix~\ref{subsec:MC_single}.  As the resultant 50\%
  detection completeness is found to depend on the input ellipticity
  $\epsilon^\mathrm{in}$, we show the 50\% detection completeness
  limits for $\epsilon^\mathrm{in} = 0.0$, 0.4, and 0.8 in the bottom
  panel of Figure~\ref{fig:size-I814}.  This simulation suggests that,
  in our ACS images, extended objects may suffer from the effect of
  limiting surface brightness if they have small ellipticities and are
  fainter than $I_{814} \sim 26$~mag, which is close to the mean
  magnitude for the ACS sources.  There are found to be a
  non-negligible fraction of the ACS sources (i.e., $9 / 54 = 16.7\%$)
  in the domain where the detection completeness limit for
  $\epsilon^\mathrm{in} = 0.0$ is below 50\%.  Hence, the number
  fraction of the LAEs having extended ACS sources can be larger than
  the observed one.

  \subsection{Ellipticity}\label{subsec:ellipticity}

  \begin{figure}
   \plotone{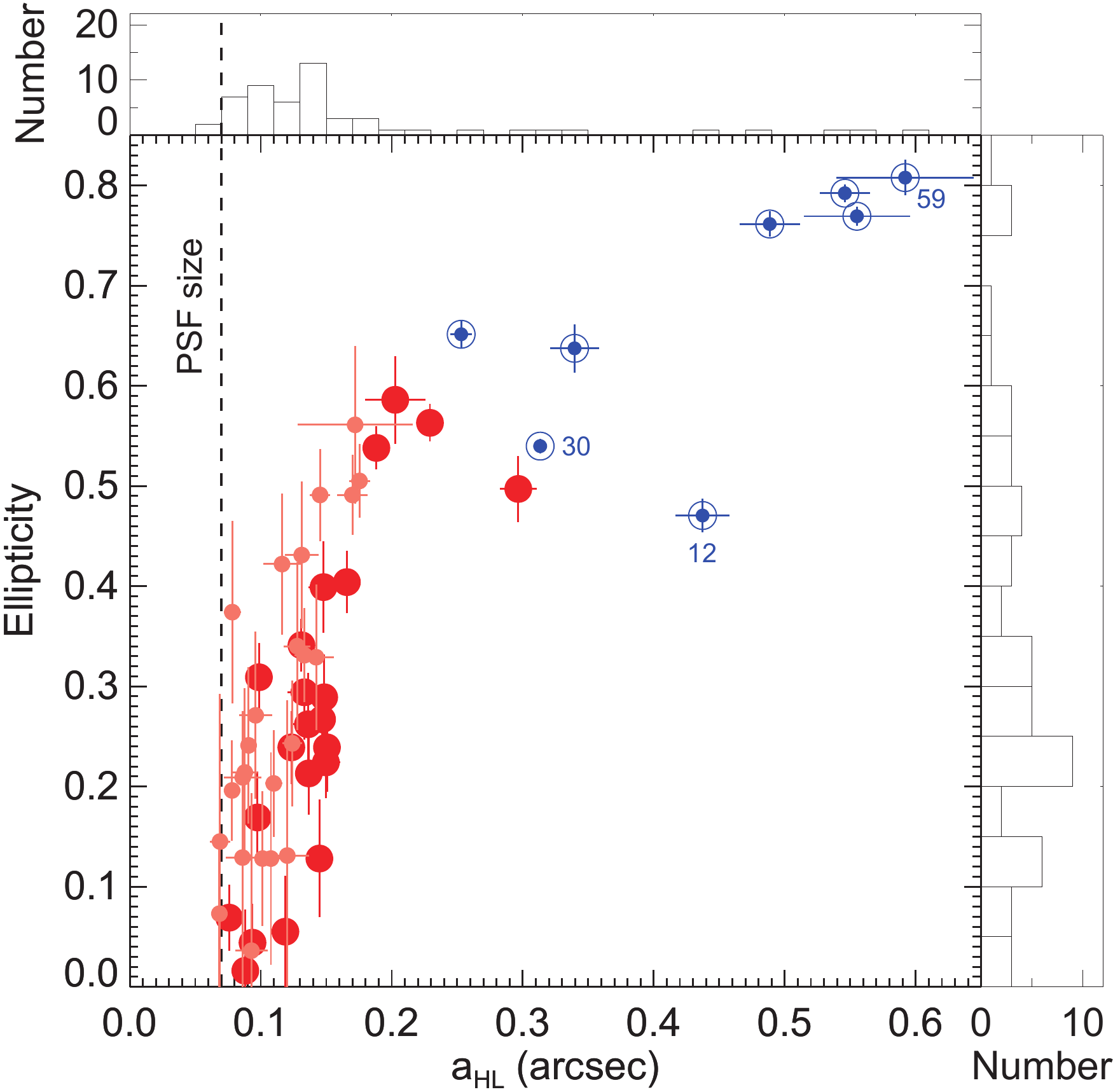}

   \caption{Distribution of the 54 ACS sources in the
   $\epsilon$--$a_\mathrm{HL}$ plane.  The symbols are the same as
   those in Figure~\ref{fig:size-I814}.  The vertical dashed line
   indicates the PSF size of the ACS F814W-band images derived from
   stars.  The ID of a double-component LAE which is an outlier in the
   positive relationship between ellipticity and $a_\mathrm{HL}$
   (i.e., \#12) is labeled for reference.  We also label the IDs of
   two double-component LAEs (i.e., \#30 and \#59) for reference.  The
   distribution in the main panel is projected onto the two side
   panels where histograms of $a_\mathrm{HL}$ and $\epsilon$ are
   displayed. \label{fig:e-aHL-total}}

  \end{figure}
  The measured ellipticities of the 54 ACS sources are widely
  distributed from 0.02 (i.e., almost round-shape) to 0.81 (i.e.,
  elongated- or ellipsoidal-shape) as shown in
  Figure~\ref{fig:e-aHL-total} and Table~\ref{tab:z4p9LAE}.  It is
  found that the double-component LAEs tend to have larger
  ellipticities than the single-component LAEs; at $\epsilon > 0.6$,
  all sources are the double-component LAEs.

  We also find a strong positive correlation between $\epsilon$ and
  $a_\mathrm{HL}$ as presented in Figure~\ref{fig:e-aHL-total} (its
  Spearman's rank-order correlation coefficient $\rho$ and Kendall's
  $\tau$ are $\rho = 0.74$ and $\tau = 0.58$, respectively); larger
  LAEs have more elongated shapes.  Moreover, all ACS sources larger
  than 0\farcs 2 have elongated morphologies ($\epsilon \gtrsim 0.5$),
  except for a double-component LAE at $a_\mathrm{HL} \sim 0\farcs 44$
  and $\epsilon \sim 0.47$ (i.e., the LAE \#12 as labeled in
  Figure~\ref{fig:e-aHL-total}).  In other words, there is no LAE with
  large size and round shape, that is, $a_\mathrm{HL} \gtrsim 0\farcs
  2$ and $\epsilon \lesssim 0.45$.  It should be emphasized that,
  since such large round-shaped galaxies can be detected if they are
  bright enough (i.e., $I_{814} \lesssim 26$~mag) as shown in
  Figure~\ref{fig:size-I814} (see also Figure~\ref{fig:MC-detecomp}),
  the absence of such galaxies can be considered as real result, not
  suffered by selection bias against them.

  It is possible that measuring the sizes and ellipticities of the
  double-component LAEs as single sources makes the correlation
  strengthen.  This is because their sizes and ellipticities are found
  to be well correlated with the separations between the two
  components (see Tables~\ref{tab:z4p9LAE} and \ref{tab:EachDouble})
  in the sense that the LAE with larger separation has larger size and
  ellipticity as a system with two components.  In the following
  section, we re-measure the sizes and ellipticities of individual ACS
  sources in the double-component LAEs separately and re-examine the
  correlation with size and ellipticity for the resultant quantities.

  \subsection{Size and Ellipticity of Individual ACS Component and
  Their Correlation}\label{subsec:double}

  As described in Section~\ref{sec:data}, the 8 ACS-detected LAEs are
  found to consist of the double components with close angular
  separation (i.e., $\lesssim 1^{\prime\prime}$) in the ACS images.
  We have shown their morphological properties measured as single
  systems with double components in the previous
  Sections~\ref{subsec:size} and \ref{subsec:ellipticity}.  However,
  the distributions of single- and double-component LAEs in both size
  and ellipticity are found to be clearly different from each other;
  the double-component LAEs have systematically larger sizes and
  ellipticities than the single-component LAEs as shown in
  Figure~\ref{fig:e-aHL-total}.  Moreover, as described in the
  previous section, the sizes and ellipticities of the
  double-component LAEs are found to be well correlated with the
  angular separation between the two components.  These findings may
  indicate that, for the 8 double-component LAEs, the morphological
  properties of individual ACS components should be measured
  separately so that they might be similar to those of the
  single-component LAEs.

  \begin{figure}
   \plotone{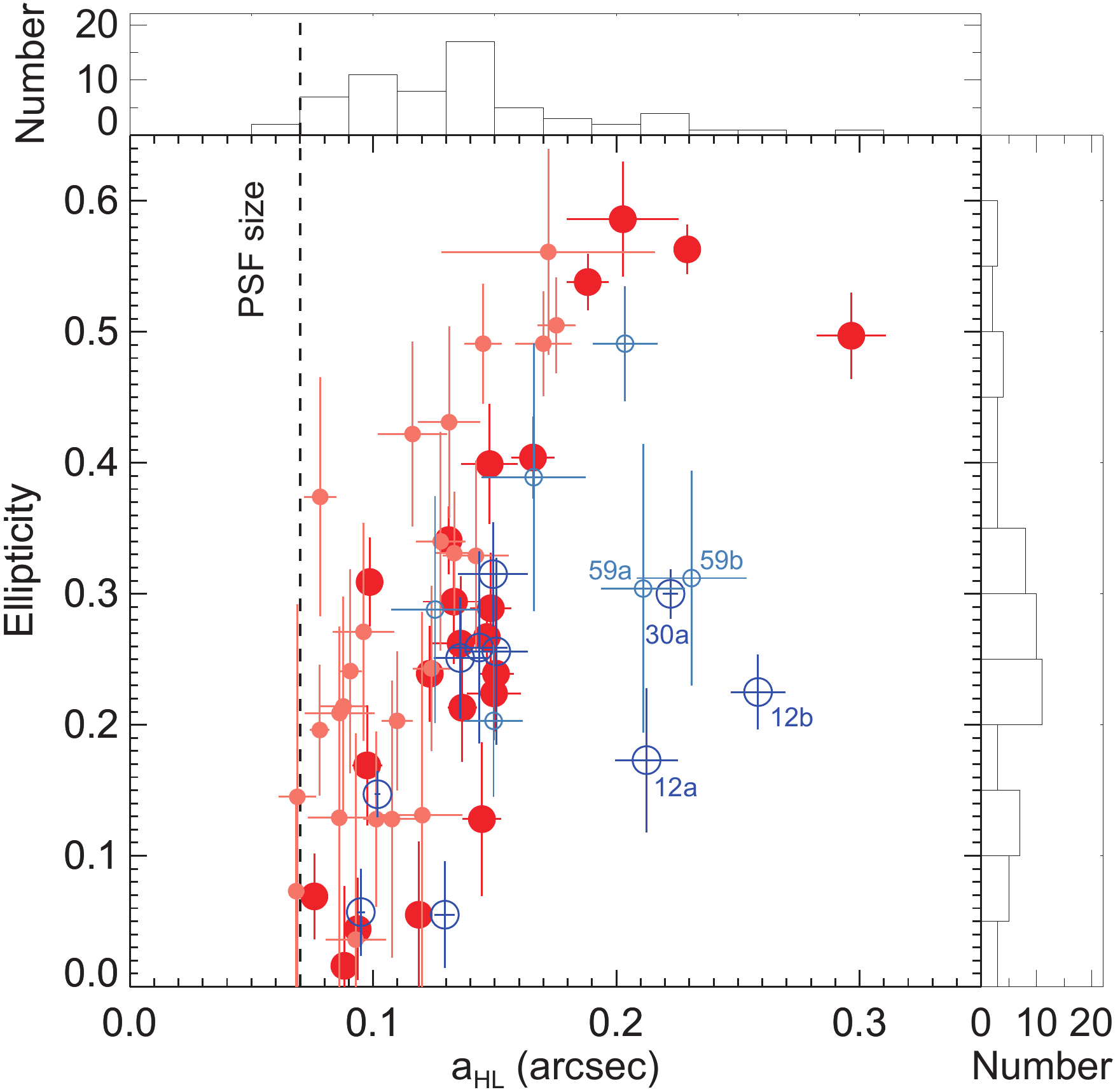}

   \caption{Same as Figure~\ref{fig:e-aHL-total} but for the
   distribution in which the double ACS components in the 8
   double-component LAEs are plotted separately.  The components of
   the single- and double-component LAEs are shown as red filled and
   blue open circles, respectively.  The double-component LAEs with
   $I_{814} \ge 26$~mag are shown by the small symbols.  The IDs of
   the 5 double-component LAEs which are outliers in the positive
   relationship between ellipticity and $a_\mathrm{HL}$ are labeled
   for reference.\label{fig:e-aHL-sep}}

  \end{figure}
  \begin{deluxetable*}{lcccccc}
   \tabletypesize{\scriptsize}
   \tablecaption{\textit{ACS} F814W-band Properties for the Individual
   Components in the 8 Double-component LAEs at $z = 4.86$ with ACS
   Data \label{tab:EachDouble}}
   \tablehead{
   \colhead{ID \# \tablenotemark{a}} &
   \colhead{$I_{814}$\tablenotemark{b}} &
   \colhead{$a_\mathrm{HL}$\tablenotemark{c}} &
   \colhead{$R_\mathrm{HL}$\tablenotemark{d}} &
   \colhead{$\epsilon (I_{814})$\tablenotemark{e}} &
   \colhead{$r_\mathrm{sep}$\tablenotemark{f}}\\
   \colhead{} &
   \colhead{(mag)} &
   \colhead{(arcsec)} &
   \colhead{(arcsec)} &
   \colhead{} &
   \colhead{(arcsec)}
   }
   \startdata
   11a & $26.18 \pm 0.08$ & $0.15 \pm 0.01$ & $0.13 \pm 0.01$ & $0.20 \pm 0.06$ & $0.59 \pm 0.01$\\
   11b & $26.37 \pm 0.10$ & $0.20 \pm 0.01$ & $0.15 \pm 0.01$ & $0.49 \pm 0.04$ & \nodata        \\
   12a & $25.43 \pm 0.07$ & $0.21 \pm 0.01$ & $0.19 \pm 0.01$ & $0.17 \pm 0.06$ & $0.47 \pm 0.01$\\
   12b & $24.49 \pm 0.03$ & $0.26 \pm 0.01$ & $0.23 \pm 0.01$ & $0.23 \pm 0.03$ & \nodata        \\
   19a & $25.86 \pm 0.07$ & $0.14 \pm 0.01$ & $0.12 \pm 0.01$ & $0.25 \pm 0.05$ & $0.70 \pm 0.01$\\
   19b & $25.76 \pm 0.07$ & $0.15 \pm 0.01$ & $0.13 \pm 0.01$ & $0.32 \pm 0.04$ & \nodata        \\
   21a & $26.72 \pm 0.12$ & $0.13 \pm 0.02$ & $0.11 \pm 0.02$ & $0.29 \pm 0.09$ & $0.98 \pm 0.01$\\
   21b & $25.32 \pm 0.04$ & $0.13 \pm 0.00$ & $0.13 \pm 0.00$ & $0.06 \pm 0.04$ & \nodata        \\
   30a & $24.56 \pm 0.02$ & $0.22 \pm 0.00$ & $0.19 \pm 0.01$ & $0.30 \pm 0.02$ & $0.36 \pm 0.00$\\
   30b & $24.69 \pm 0.01$ & $0.10 \pm 0.00$ & $0.09 \pm 0.00$ & $0.15 \pm 0.02$ & \nodata        \\
   37a & $25.94 \pm 0.09$ & $0.15 \pm 0.01$ & $0.13 \pm 0.01$ & $0.26 \pm 0.07$ & $0.41 \pm 0.01$\\
   37b & $25.98 \pm 0.10$ & $0.14 \pm 0.01$ & $0.13 \pm 0.01$ & $0.26 \pm 0.07$ & \nodata        \\
   59a & $26.39 \pm 0.11$ & $0.21 \pm 0.02$ & $0.18 \pm 0.02$ & $0.30 \pm 0.11$ & $0.53 \pm 0.01$\\
   59b & $26.08 \pm 0.10$ & $0.23 \pm 0.02$ & $0.19 \pm 0.02$ & $0.31 \pm 0.08$ & \nodata        \\
   68a & $25.17 \pm 0.02$ & $0.09 \pm 0.00$ & $0.09 \pm 0.00$ & $0.06 \pm 0.03$ & $0.97 \pm 0.01$\\
   68b & $27.23 \pm 0.15$ & $0.17 \pm 0.02$ & $0.13 \pm 0.02$ & $0.39 \pm 0.10$ & \nodata
   \enddata

   \tablecomments{(a) The LAE ID given in Shioya et al.~(2009).  (b)
   SExtractor's MAG\_AUTO magnitude and its $1\sigma$ error.  (c)
   Half-light major radius and its $1\sigma$ error measured on ACS
   F814W-band images.  (d) Half-light radius and its $1\sigma$ error
   measured on ACS F814W-band images.  (e) Ellipticity and its
   $1\sigma$ error measured on ACS F814W-band images.  (f) Angular
   separation between the two components in a double-component LAE and
   its $1\sigma$ error.}

  \end{deluxetable*}
  Figure~\ref{fig:e-aHL-sep} shows the resultant distribution in
  which, even for the double-component LAEs, both of $\epsilon$ and
  $a_\mathrm{HL}$ of the individual ACS components are measured
  separately using SExtractor with the same parameters shown in
  Table~\ref{table:SExParam}.  The morphological properties for the
  individual components of the 8 double-component LAEs as well as
  their angular separations are listed in Table~\ref{tab:EachDouble}.
  Compared with the distributions shown in
  Figure~\ref{fig:e-aHL-total}, the distribution of the
  double-component LAEs becomes similar to that of the
  single-component LAEs, while there seem to be five outliers at
  $a_\mathrm{HL} \sim 0\farcs 21$--0\farcs 26 and $\epsilon \sim
  0.17$--0.31; the outliers are the LAEs \#12a, \#12b, \#30a, \#59a,
  and \#59b as labeled in Figure~\ref{fig:e-aHL-sep}.  As shown in
  Figure~\ref{fig:ThumbnailsDouble} and Table~\ref{tab:EachDouble},
  the morphological properties of these outliers may be affected by
  the other component of a pair because of the close angular
  separation $r_\mathrm{sep}$, which is characterized by
  $r_\mathrm{sep} \lesssim 2.5a_\mathrm{HL}$.  On the other hand,
  those of other double-component LAEs are found to be characterized
  by $r_\mathrm{sep} \gtrsim 3a_\mathrm{HL}$ and hence they could not
  be affected by the other component.

  Figure~\ref{fig:e-aHL-sep} also shows that the positive correlation
  between $\epsilon$ and $a_\mathrm{HL}$ still exists, while it
  becomes weaker ($\rho = 0.64$ and $\tau = 0.46$) compared with the
  correlation shown in Figure~\ref{fig:e-aHL-total} ($\rho = 0.74$ and
  $\tau = 0.58$).  If we consider that, as usually do, the LAE
  consists of thin disk and the ellipticity of ACS source reflects the
  inclination angle to its disk, the existence of such correlation and
  the absence of the sources with large $a_\mathrm{HL}$ and small
  $\epsilon$ are unnatural.  We will discuss their origin(s) in
  Section~\ref{subsec:dis-mor}.

 \section{DISCUSSION}

  \subsection{Comparison of Size and
  Ellipticity with Those in The Literatures}\label{subsec:comparison}

  As shown in Section~\ref{subsec:size}, the single-component LAEs are
  found to be widely distributed in $a_\mathrm{HL}$ of rest-frame UV
  continuum from 0\farcs 07 to 0\farcs 30 ($\langle a_\mathrm{HL}
  \rangle = 0\farcs 13$), which correspond to the physical sizes of
  0.45~kpc and 1.90~kpc (0.83~kpc for the mean) at $z = 4.86$.  These
  measured sizes are quantitatively consistent with the previous
  measurements for the sizes in rest-frame UV continuum of the LAEs at
  $z \sim 2$--6 compiled in Malhotra et al. (2012; see also Hagen et
  al. 2014 for more recent observational results of the size
  measurements for the LAEs at $z = 1.9$--3.6).  Therefore, our size
  measurements for the LAEs at $z = 4.86$ provide a further support to
  the result of Malhotra et al. (2012), that is, the sizes of LAEs in
  rest-frame UV continuum do not show redshift evolution in $z \sim
  2$--6.

  In contrast, the sizes of the double-component LAEs are
  systematically larger than those of the previous measurements;
  $a_\mathrm{HL}$ of the double-component LAEs ranges from 0\farcs 25
  to 0\farcs 59 with a mean of 0\farcs 44 (see
  Figure~\ref{fig:e-aHL-total}).  On the other hand, their sizes are
  found to be also consistent with those in the literature if their
  sizes of the individual ACS components are adopted; as shown in
  Table~\ref{tab:EachDouble}, $a_\mathrm{HL}$ of the individual
  components in the double-component LAEs are in the range of 0\farcs
  09--0\farcs 26 with a mean of 0\farcs 17 (see
  Figure~\ref{fig:e-aHL-sep}).  Therefore, it seems to be more natural
  that the individual ACS components in the double-component LAEs are
  typical LAEs and that the double-component LAEs are interacting
  and/or merging galaxies compared to the interpretation that they are
  sub-components (e.g., star-forming clumps) in an LAE.  We will
  discuss these interpretations of our ACS sources further in
  Section~\ref{subsec:Implication}.

  In terms of $\epsilon$, we also presented in
  Figure~\ref{fig:e-aHL-sep} (i.e., the case that the ACS components
  in the double-component LAEs are treated separately) that the
  distribution of the ACS sources in $\epsilon$ shows a peak around
  the mean ellipticity of $\langle \epsilon \rangle = 0.27$ and has
  long tails toward both smaller and larger ellipticities in the range
  of $\epsilon = 0.02$--0.59.  This distribution in $\epsilon$ is
  found to be quite similar to the previous observational estimates
  for the LAEs at $z \sim 2.2$ (Shibuya et al. 2014) and $z \sim 3.1$
  (Gronwall et al. 2011).  Although we found the positive correlation
  between $\epsilon$ and $a_\mathrm{HL}$ as shown in
  Figure~\ref{fig:e-aHL-sep}, such correlation has not been
  investigated so far; therefore, we do not have any previous results
  that can be compared with ours.  It is still unclear whether or not
  such correlation is seen among LAEs at different redshifts and how
  it evolves with redshift.  Nevertheless, we will discuss the
  origin(s) of the positive correlation in
  Section~\ref{subsec:dis-mor}.

  \subsection{Implication for the Sizes of the ACS-undetected LAEs}

  \begin{figure}
   \plotone{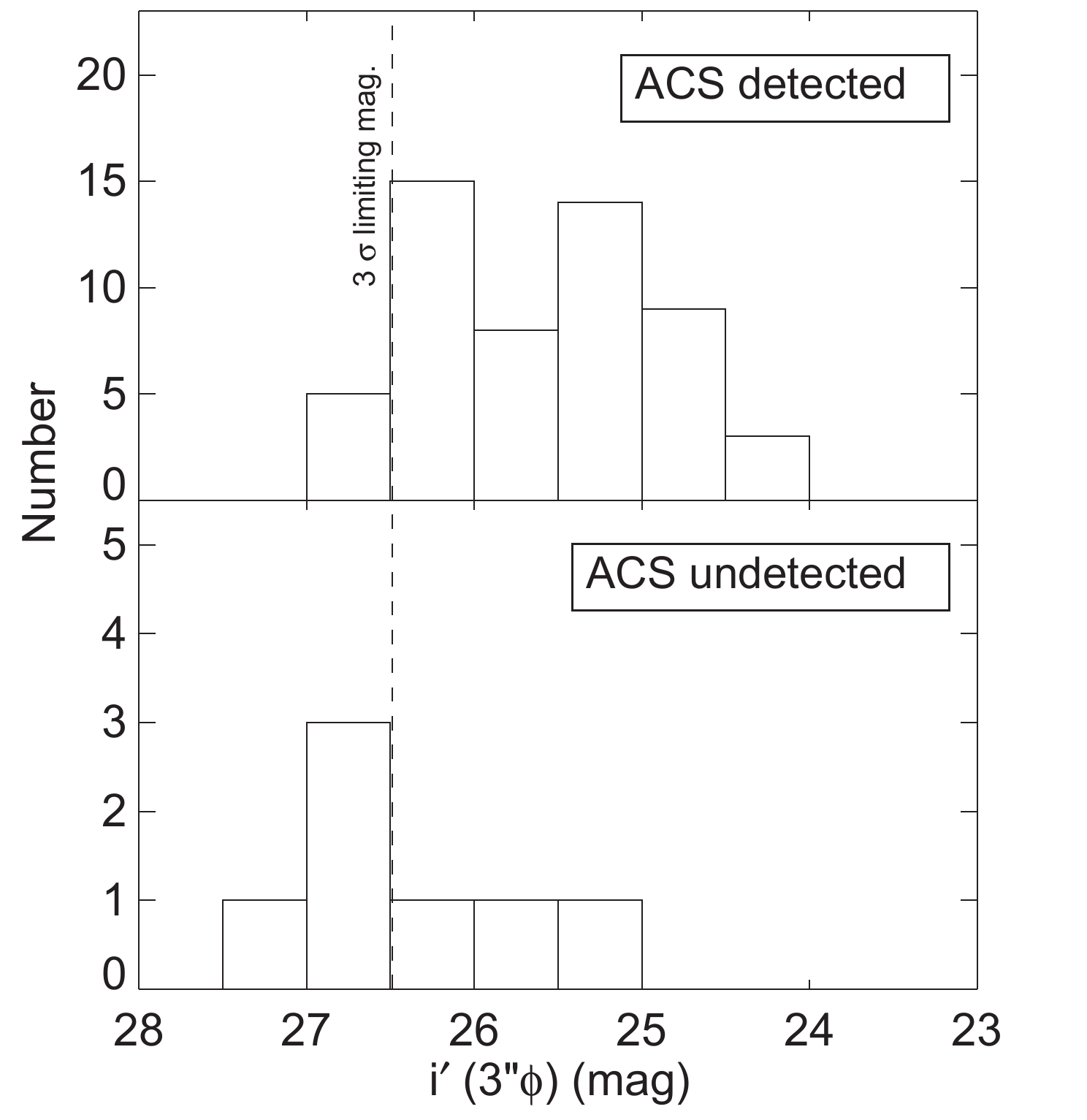}

   \caption{Frequency distributions of $i^\prime$--band magnitudes for
   the 54 ACS-detected (\textit{top}) and 7 ACS-undetected LAEs
   (\textit{bottom}).  The $3\sigma$ limiting magnitude of
   $i^\prime$--band images in a $3^{\prime\prime}$ diameter aperture,
   26.49~mag, is also represented by the vertical dashed
   line. \label{fig:ipDist}}

  \end{figure}
  Among the 61 LAEs with the ACS F814W-band imaging data, 7 LAEs are
  not detected in the ACS images.  Here we try to estimate the
  half-light radii of these ACS-undetected LAEs using the correlation
  between $I_{814}$ and $i^\prime$ found for the ACS-detected LAEs
  (see Figure~\ref{fig:I814vsNB711}) and the $i^\prime$--band magnitude
  distribution of the ACS-undetected LAEs.

  As shown in Figure~\ref{fig:ipDist}, while the ACS-undetected LAEs
  are found to be at fainter part in the $i^\prime$--band magnitude
  distribution compared with the ACS-detected LAEs, most of the
  ACS-undetected LAEs have similar $i^\prime$--band magnitudes to those
  of the ACS-detected LAEs.  Considering the result of $I_{814}
  \approx i^\prime$ found for the ACS-detected LAEs, the
  ACS-undetected LAEs with similar $i^\prime$--band magnitudes to the
  ACS-detected LAEs ought to be detected if they are compact and have
  small $R_\mathrm{HL}$.  Therefore, the results of their
  non-detection in the ACS images imply that the surface brightnesses
  of the 7 ACS-undetected LAEs are too low to be detected; that is,
  even if they are bright enough to be detected in $I_{814}$, they
  cannot be detected in ACS image in the case that they are spatially
  extended significantly as discussed in Section~\ref{subsec:size}
  (see the 50\% detection completeness shown in the top panel of
  Figure~\ref{fig:size-I814}).  Therefore, large $R_\mathrm{HL}$ can
  be expected for the ACS-undetected LAEs.

  We can estimate the half-light radii of the ACS-undetected LAEs as
  follows.  First, we evaluate the expected $I_{814}$-band magnitude
  from $i^\prime$--band magnitude, $I_{814} (i^\prime)$, using the
  best-fit linear relation between $I_{814}$- and $i^\prime$--band
  magnitudes for the ACS-detected LAEs: $I_{814}(i^\prime) = 0.88
  (i^\prime - 29.0) + 29.0$.  This is motivated by the result that
  $I_{814}$ is well correlated with $i^\prime$ as described in
  Section~\ref{subsec:ACS}.  Providing $i^\prime = 25.23$--27.04~mag
  for the ACS-undetected LAEs, we obtain $I_{814}(i^\prime) =
  25.7$--27.3~mag.  Then, as the ACS-undetected LAEs are expected to
  be in the domain on the $R_\mathrm{HL}$--$I_{814}$ plane, where
  detection completeness is low, a lower-limit of $R_\mathrm{HL}$ for
  the ACS-undetected LAEs can be estimated from $I_{814}(i^\prime)$
  and the curve in the $R_\mathrm{HL}$--$I_{814}$ plane at which
  detection completeness for exponential disk objects with the input
  ellipticity of $\epsilon^\mathrm{in} = 0.0$ is 50\% (see
  Figure~\ref{fig:size-I814}).  As a result, the expected half-light
  radii of the ACS-undetected LAEs are $R_\mathrm{HL} \gtrsim 0\farcs
  07$--0\farcs 32.  This result may imply that there are some LAEs
  with very large $R_\mathrm{HL}$ among the ACS-undetected LAEs.

  \subsection{Comparison of the Sizes in Ly$\alpha$ and UV Continuum}

  \begin{figure}
   \plotone{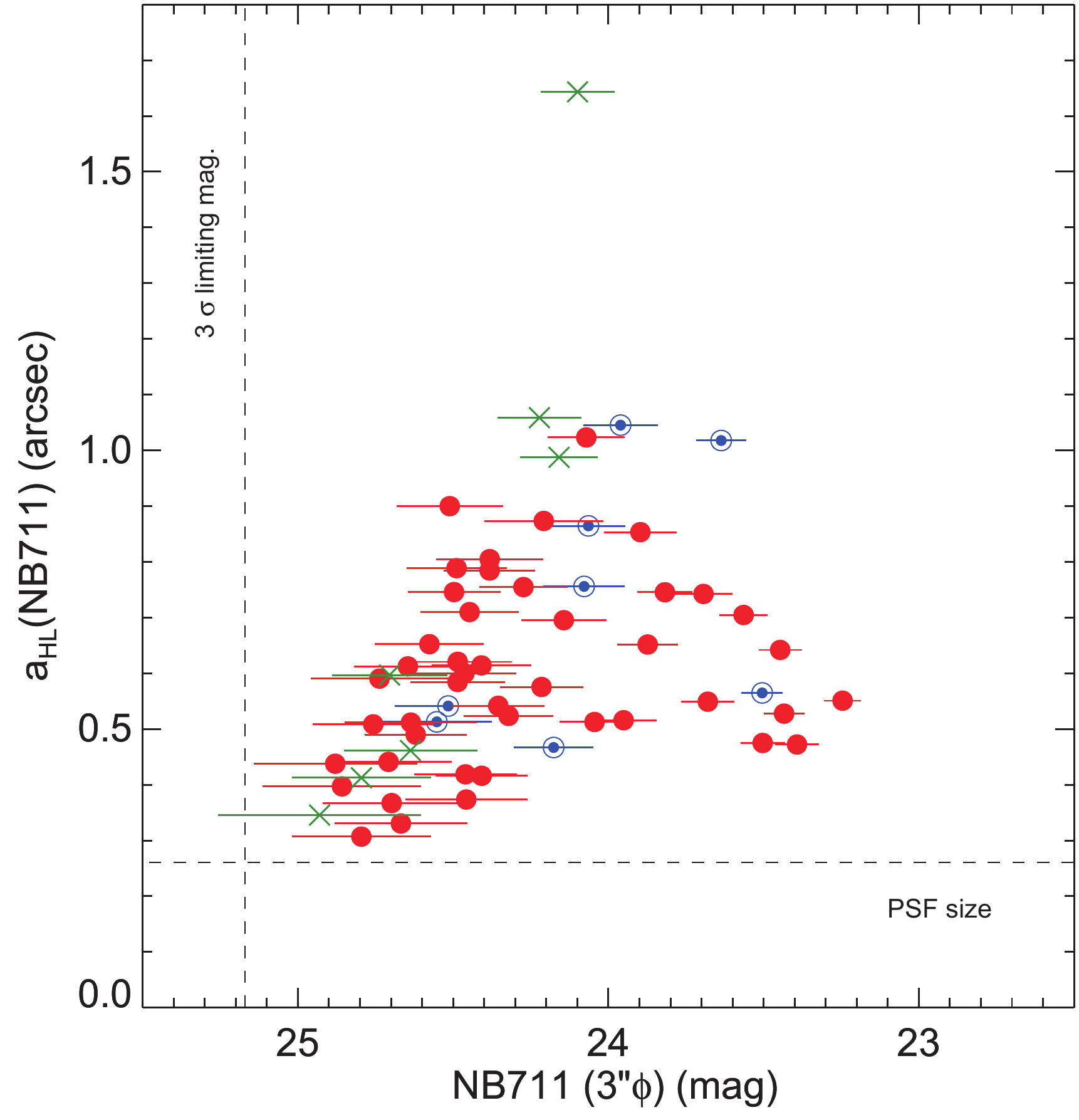}

   \caption{Distribution of the 60 LAEs with ACS F814W--band imaging
   data and with measured $a_\mathrm{HL} (\mathrm{NB711})$ in the
   $a_\mathrm{HL} (\mathrm{NB711})$--$NB711$ plane.  The 7
   ACS-undetected LAEs are shown by green crosses and the 45 (8)
   ACS-detected LAEs with single (double) component(s) are represented
   by red filled (blue double) circles.  The $3\sigma$ limiting
   magnitude in a $3^{\prime\prime}$ diameter aperture and the PSF
   half-light radius of NB711--band images, 25.17~mag and 0\farcs 25,
   respectively, are also indicated by the vertical dotted and
   horizontal dashed lines.  \label{fig:FWHMvsNB711}}

  \end{figure}
\begin{figure}
   \plotone{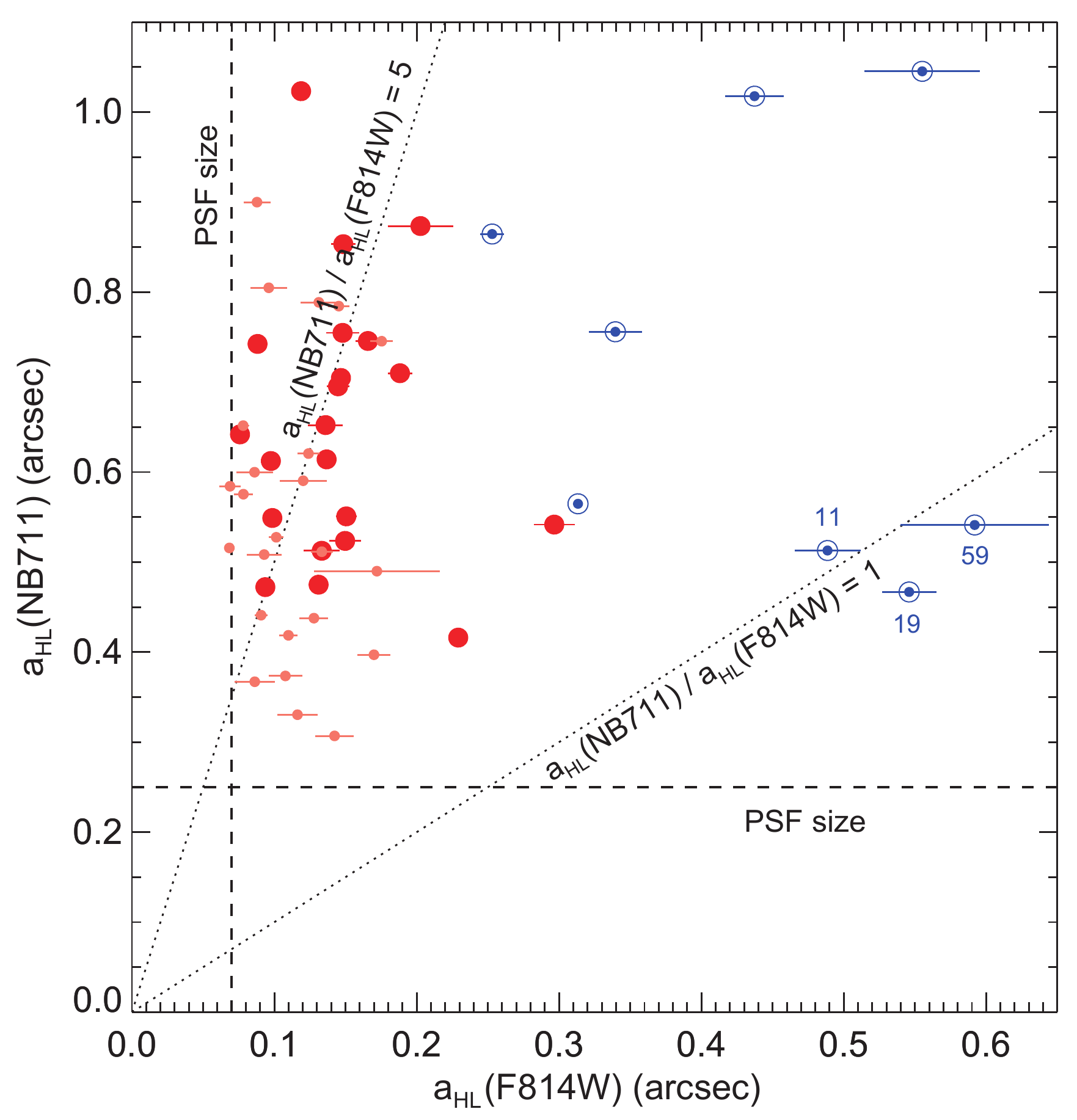}

   \caption{Distribution of the 53 ACS sources with measured
   $a_\mathrm{HL} (\mathrm{NB711})$ in the $a_\mathrm{HL}
   \mathrm{(NB711)}$--$a_\mathrm{HL} \mathrm{(F814W)}$ plane.  Symbols
   are the same as those in Figure~\ref{fig:e-aHL-total}.  The PSF
   sizes of ACS F814W-- and NB711--band images are presented by
   vertical and horizontal dashed lines, respectively.  The dotted
   lines represent $a_\mathrm{HL} \mathrm{(NB711)} / a_\mathrm{HL}
   \mathrm{(F814W)} = 1$ (lower) and 5 (upper) as labeled.  The IDs of
   the 3 outliers located at $a_\mathrm{HL} (\mathrm{NB711}) /
   a_\mathrm{HL} (\mathrm{F814W}) \approx 1$ are also labeled for
   reference.  \label{fig:RHLvsFWHM}}

  \end{figure}

  In Section~\ref{subsec:size}, we found that the ACS-detected LAEs
  have a wide range of $a_\mathrm{HL}$ in the ACS F814W-band images
  from 0\farcs 07 to 0\farcs 59 ($\langle a_\mathrm{HL}
  (\mathrm{F814W}) \rangle = 0\farcs 175 \pm 0\farcs
  017$\footnote{Only in this Section and Figure~\ref{fig:RHLvsFWHM},
  in order to avoid confusion with $a_\mathrm{HL} (\mathrm{NB711})$,
  we refer the half-light major radius in the ACS F814--band image as
  $a_\mathrm{HL} (\mathrm{F814W})$ rather than $a_\mathrm{HL}$ used in
  the other parts of this paper.}).  These angular scales correspond
  to the physical scales of 0.45--3.8~kpc ($1.11\pm 0.11$~kpc for the
  mean) at $z = 4.86$.  As shown in Figure~\ref{fig:FWHMvsNB711}, the
  half-light major radii in the NB711--band images, $a_\mathrm{HL}
  (\mathrm{NB711})$, of the 61 LAEs with ACS data are also found to
  widely distribute in 0\farcs 31--1\farcs 64, corresponding to the
  physical scales of 1.97--10.45~kpc at $z = 4.86$.  Since the PSF
  half-light radius of NB711--band images is 0\farcs 25\footnote{Note
  that the PSF FWHM of the NB711--band images is estimated to be
  0\farcs 79 (Shioya et al. 2009).}, most LAEs are significantly
  extended in Ly$\alpha$ emissions.  This result is consistent with
  previous studies (e.g., Taniguchi et al. 2005, 2009, 2015; Malhotra
  et al. 2012; Mawatari et al. 2012; Momose et al. 2014).

  It is interesting to examine the relation between $a_\mathrm{HL}
  (\mathrm{NB711})$ and $a_\mathrm{HL} (\mathrm{F814W})$ for the
  ACS-detected LAEs.  As shown in Figure~\ref{fig:RHLvsFWHM},
  $a_\mathrm{HL} (\mathrm{NB711})$ is systematically larger than
  $a_\mathrm{HL} (\mathrm{F814W})$ except for three double-component
  LAEs with $a_\mathrm{HL} (\mathrm{NB711}) / a_\mathrm{HL}
  (\mathrm{F814W}) \approx 1$ (i.e., the LAEs \#11, \#19, and \#59
  have the ratios of 1.04, 0.85, and 0.96, respectively).  The ratio
  of $a_\mathrm{HL} \mathrm{(NB711)} / a_\mathrm{HL} (\mathrm{F814W})$
  is widely distributed from $\approx 1$ to $\approx 10$.  This result
  may be a consequence of the non-detection of extended UV continuum
  in the ACS images since extended sources are difficult to be
  detected as discussed in Section~\ref{subsec:size}.  Nevertheless,
  the large ratio of $a_\mathrm{HL} \mathrm{(NB711)} / a_\mathrm{HL}
  (\mathrm{F814W})$ can be a real feature for high-$z$ LAEs.  In this
  case, it is suggested that the compact star-forming regions in the
  LAEs (i.e., $\lesssim 0\farcs 30$ or $\lesssim 1.9$~kpc) observed by
  the ACS F814W-band images ionize the surrounding gas, which emits
  spatially extended Ly$\alpha$ (i.e., $\gtrsim 0\farcs 3$ or $\gtrsim
  1.9$~kpc) detected in the NB711-band images.

  \subsection{Origin of the Correlation between Ellipticity and
  Size}\label{subsec:dis-mor}

  As described in Section~\ref{subsec:ellipticity}, the ACS sources
  show a strong positive correlation between ellipticity $\epsilon$
  and half-light major radius $a_\mathrm{HL}$, that is, larger ACS
  sources have more elongated morphologies
  (Figure~\ref{fig:e-aHL-total}).  As shown in
  Section~\ref{subsec:double} and Figure~\ref{fig:e-aHL-sep}, while
  the correlation is found to be strengthened by our measurements of
  $\epsilon$ and $a_\mathrm{HL}$ for the two ACS sources in individual
  double-component LAEs collectively, the correlation still exists if
  $\epsilon$ and $a_\mathrm{HL}$ are measured for the two ACS sources
  in double-component LAEs separately.

  \begin{figure*}
   \plotone{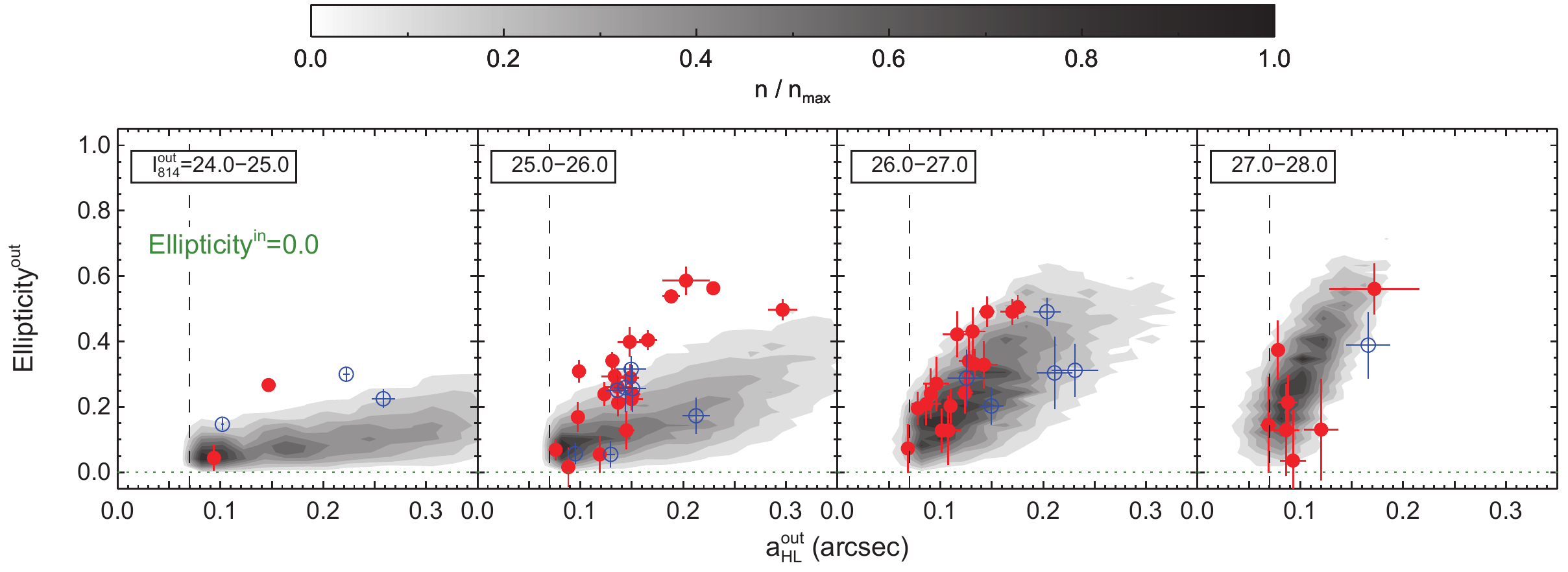}
   \plotone{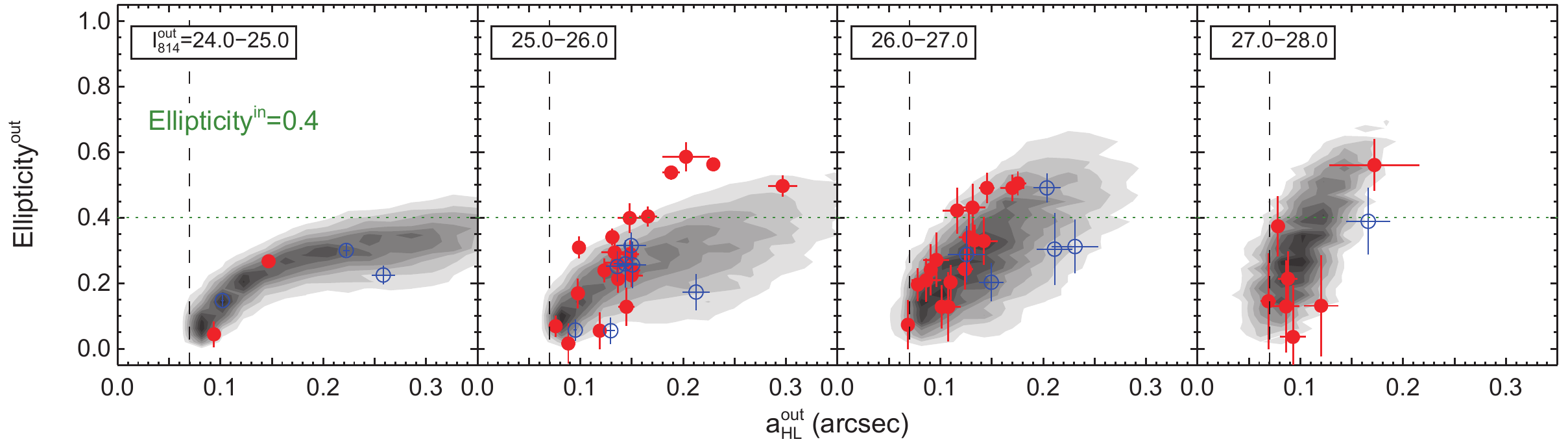}
   \plotone{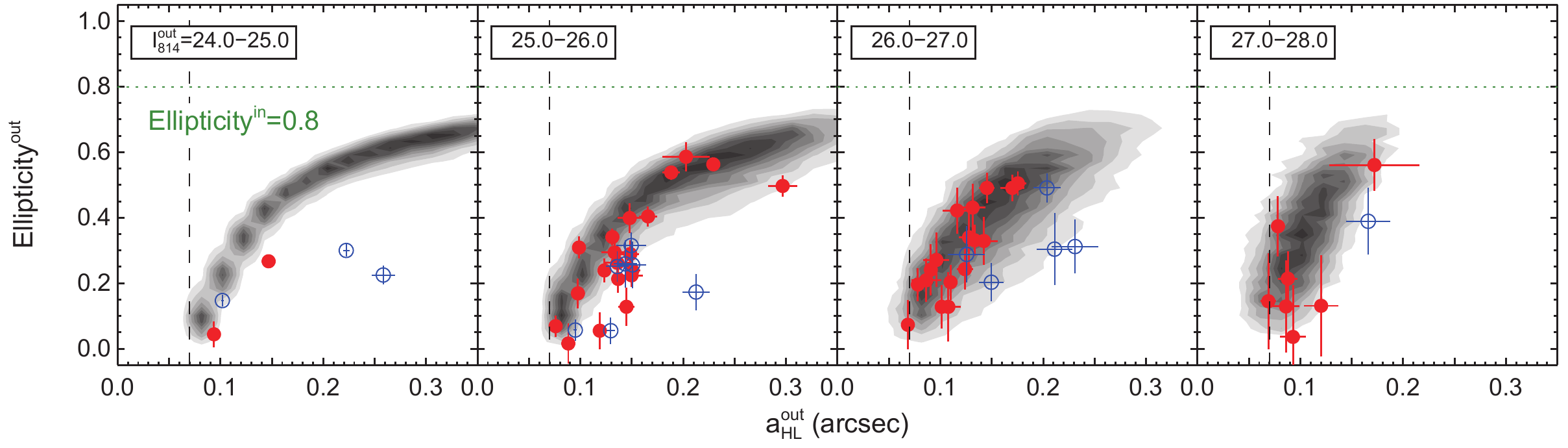}

   \caption{Distributions of the artificial sources in the
   $\epsilon$--$a_\mathrm{HL}$ plane.  1000 artificial sources with
   the exponential light profile, each of which is assumed to consist
   of a single component, are generated by a Monte Carlo simulation
   and prepared for each set of input parameters, that is,
   $I_{814}^\mathrm{in} = 24.0$--28.0~mag, $a_\mathrm{HL}^\mathrm{in}
   = 0\farcs 03$--0\farcs 36, and $\epsilon^\mathrm{in} = 0.0$--0.9
   (see Appendix~\ref{subsec:MC_single} for details).  The resultant
   distributions are plotted separately for the artificial sources
   with the input ellipticities $\epsilon^\mathrm{in}$ of 0.0 (top),
   0.4 (middle), and 0.8 (bottom), as well as those with the output
   magnitudes $I_{814}^\mathrm{out}$ as labeled in each panel.  The
   grayscale represents the number of the artificial sources in the
   linear scale from 0 (white) to $n_\mathrm{max}$ (black), where
   $n_\mathrm{max}$ is the maximum number of the artificial sources in
   a grid and different among panels.  Note that these grayscales are
   evaluated only from the detected artificial sources.  The vertical
   dashed lines indicate the PSF half-light radius $R_\mathrm{PSF}$
   and the horizontal dotted lines represent $\epsilon^\mathrm{in}$.
   The observed 62 ACS sources are overlaid with the same symbols as
   those in Figure~\ref{fig:e-aHL-sep}.  \label{fig:MCsingle}}

  \end{figure*}
  Here, we examine the possibilities that the observed correlation
  between $\epsilon$ and $a_\mathrm{HL}$ is (1) an
  ``\textit{apparent}'' correlation caused by deformation effects for
  a single source (e.g., pixelization, PSF broadening, and shot noise)
  and (2) the ``\textit{intrinsic}'' correlation originated from
  blending with unresolved double or multiple sources through Monte
  Carlo simulations.  We consider the correlation between $\epsilon$
  and $a_\mathrm{HL}$ for the individual ACS sources of the
  double-component LAEs (i.e., the correlation shown in
  Figure~\ref{fig:e-aHL-sep}) since it is more natural interpretation
  as described in Section~\ref{subsec:double}.

   \subsubsection{Apparent Correlation Caused by Deformation
   Effects}\label{subsubsec:SingleDeform}

   As shown in the previous section, our ACS sources are typically
   very compact and faint.  In general, the sizes and ellipticities of
   compact sources whose angular scales are comparable to the pixel
   scale can be modified because of the pixelization of the digital
   images depending on their places on the pixels.  However, since the
   ACS images we used have sufficiently large PSF half-light radius of
   $R_\mathrm{PSF} = 0\farcs 07$ compared to the pixel scale of
   0\farcs 03, such pixelization seems not to affect the morphological
   parameters of the detected sources, which are definitely affected
   by PSF broadening.  The sizes and ellipticities of faint sources
   whose surface brightnesses are comparable to the surface brightness
   limit of imaging data also tend to be modified by shot noise; that
   is, if a bright pixel contaminated by shot noise appears near a
   compact faint source, they could be blended with each other and
   detected as a single elongated source via our source detection
   using SExtractor.  The angular separation between the
   noise-contaminated pixel and source is translated into the size and
   ellipticity of the detected source.  Therefore, a positive
   correlation between $\epsilon$ and $a_\mathrm{HL}$ is naturally
   expected to emerge even if source has intrinsically compact and
   perfectly round shape.

   \begin{figure*}
    \plotone{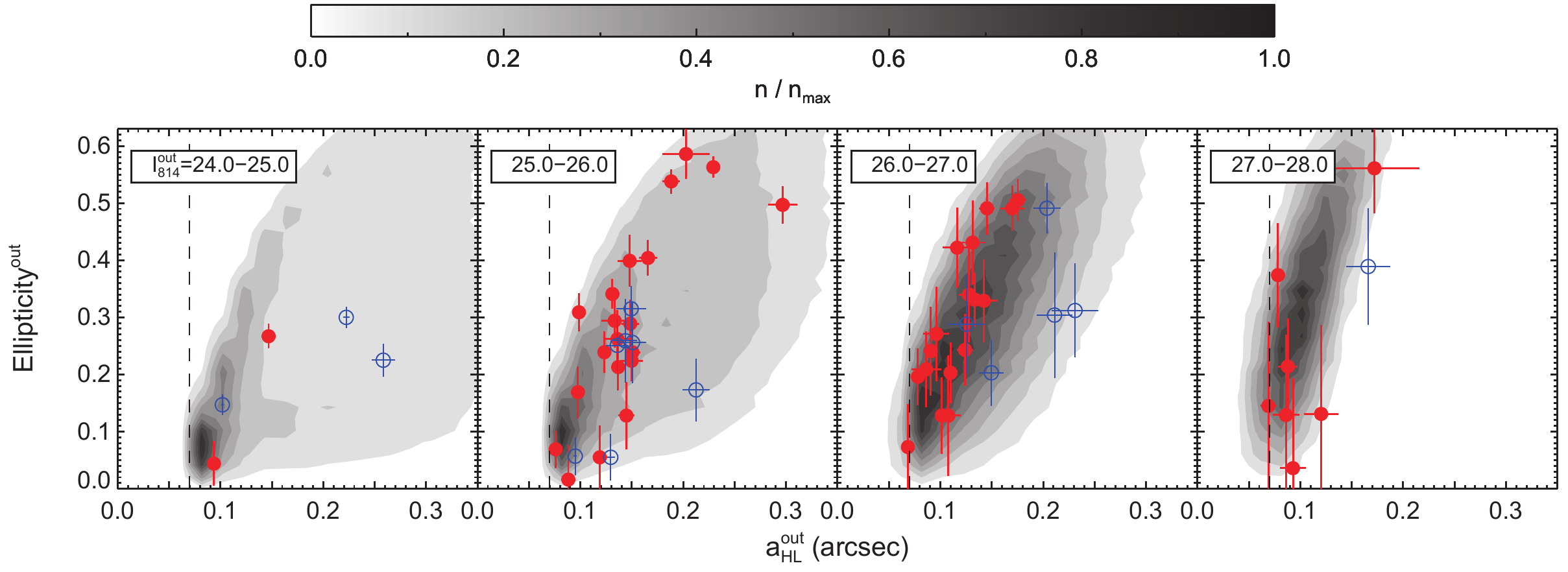}

    \caption{Same as Figure~\ref{fig:MCsingle}, but for the result of
    a Monte Carlo simulation for the artificial sources with various
    $\epsilon^\mathrm{in}$ uniformly distributed in
    0.0--0.9. \label{fig:MCsingle_all}}

   \end{figure*}
   In order to examine these deformation effects from PSF broadening
   and shot noise on the distribution in the
   $\epsilon$--$a_\mathrm{HL}$ plane, we perform the Monte Carlo
   simulation which is the same as the one done to estimate the
   detection completeness (see Appendix~\ref{subsec:MC_single} for the
   details).  Figure~\ref{fig:MCsingle} shows the resultant
   distributions of the detected artificial sources with
   $\epsilon^\mathrm{in} = 0.0$ (top), 0.4 (middle), and 0.8 (bottom)
   in the $\epsilon^\mathrm{out}$--$a_\mathrm{HL}^\mathrm{out}$ plane.
   As expected, the deformation effects are found to produce a
   correlation between $\epsilon$ and $a_\mathrm{HL}$ which is similar
   to the observed correlation, although the sources intrinsically
   distribute in the
   $\epsilon^\mathrm{in}$--$a_\mathrm{HL}^\mathrm{in}$ plane
   uniformly.

   How do these deformation effects produce such an apparent
   correlation between $\epsilon$ and $a_\mathrm{HL}$?  The PSF
   broadening significantly affects the shapes of the detected sources
   with small sizes, in the sense that their measured ellipticities
   $\epsilon^\mathrm{out}$ converge on the PSF ellipticity,
   $\epsilon_\mathrm{PSF} \approx 0$, regardless of
   $I_{814}^\mathrm{in}$ and $\epsilon^\mathrm{in}$ of them.  Since
   this effect becomes less important for larger sources with
   sufficiently bright surface brightnesses, their
   $\epsilon^\mathrm{out}$ are expected to be reproduced as
   $\epsilon^\mathrm{out} \sim \epsilon^\mathrm{in}$.  Therefore, if
   the sources have non-zero $\epsilon^\mathrm{in}$, a correlation
   between $\epsilon^\mathrm{out}$ and $a_\mathrm{HL}^\mathrm{out}$
   emerges, as shown in the two left-most panels for
   $\epsilon^\mathrm{in} = 0.4$ and 0.8 of Figure~\ref{fig:MCsingle}.
   On the other hand, the effects of shot noise can be easily seen in
   the apparent positive correlation between $\epsilon^\mathrm{out}$
   and $a_\mathrm{HL}^\mathrm{out}$ for the sources with
   $\epsilon^\mathrm{in} = 0.0$; the ellipticities of the detected
   sources increase with $a_\mathrm{HL}^\mathrm{out}$ although their
   input ellipticities are exactly zero.  The emergence of the
   correlation can be interpreted via a combination of lower detection
   completenesses and larger influences of noise-contaminated pixel
   for the sources with lower surface brightnesses, that is, those
   with larger sizes and/or smaller ellipticities (see
   Figure~\ref{fig:MC-detecomp}).  These effects are more significant
   for the sources with fainter magnitudes of $I_{814}^\mathrm{out}$
   and hence the slopes of the correlation become steeper for fainter
   sources.  For the sources with $I_{814}^\mathrm{out} > 26$~mag,
   since the effects of shot noise are dominant, the correlation does
   not depend on $\epsilon^\mathrm{in}$ significantly as shown in the
   two right-most panels of Figure~\ref{fig:MCsingle}.

   As a combination of these deformation effects, the detected
   artificial sources distribute similar to the observed distribution
   in the $\epsilon^\mathrm{out}$--$a_\mathrm{HL}^\mathrm{out}$ plane
   as shown in Figure~\ref{fig:MCsingle_all}, although they are
   uniformly distributed in the
   $\epsilon^\mathrm{in}$--$a_\mathrm{HL}^\mathrm{in}$ plane.  This
   result may indicate that the observed correlation between
   $\epsilon$ and $a_\mathrm{HL}$ is apparent one caused by the
   deformation effects.  The dispersion of $\epsilon^\mathrm{out}$ for
   a given $a_\mathrm{HL}^\mathrm{out}$ is predicted to be larger for
   the sources with bright magnitudes of $I_{814}^\mathrm{out}$.  This
   is because the distributions of the brighter sources in the
   $\epsilon^\mathrm{out}$--$a_\mathrm{HL}^\mathrm{out}$ plane do
   depend on $\epsilon^\mathrm{in}$ and those of the fainter sources
   do not.  Our simulation suggests that, in order to reproduce the
   distributions of the LAEs with relatively bright (i.e., $I_{814}
   \approx 25$--26~mag) and large sizes and ellipticities (i.e.,
   $a_\mathrm{HL} \approx 0\farcs 20$ and $\epsilon \approx 0.45$),
   intrinsically large ellipticities (i.e., $\epsilon^\mathrm{in}
   \gtrsim 0.8$) are required.  We note that the observed distribution
   can be reproduced even better if the artificial sources have a
   Gaussian distribution peaked at $(a_\mathrm{HL}^\mathrm{in},\
   \epsilon^\mathrm{in}) \sim (0\farcs 15,\ 0.3)$.

   \subsubsection{Intrinsic Correlation Originated from Blending with
   Unresolved Double/Multiple Sources}\label{subsubsec:DoubleDeform}

   \begin{figure*}
    \plotone{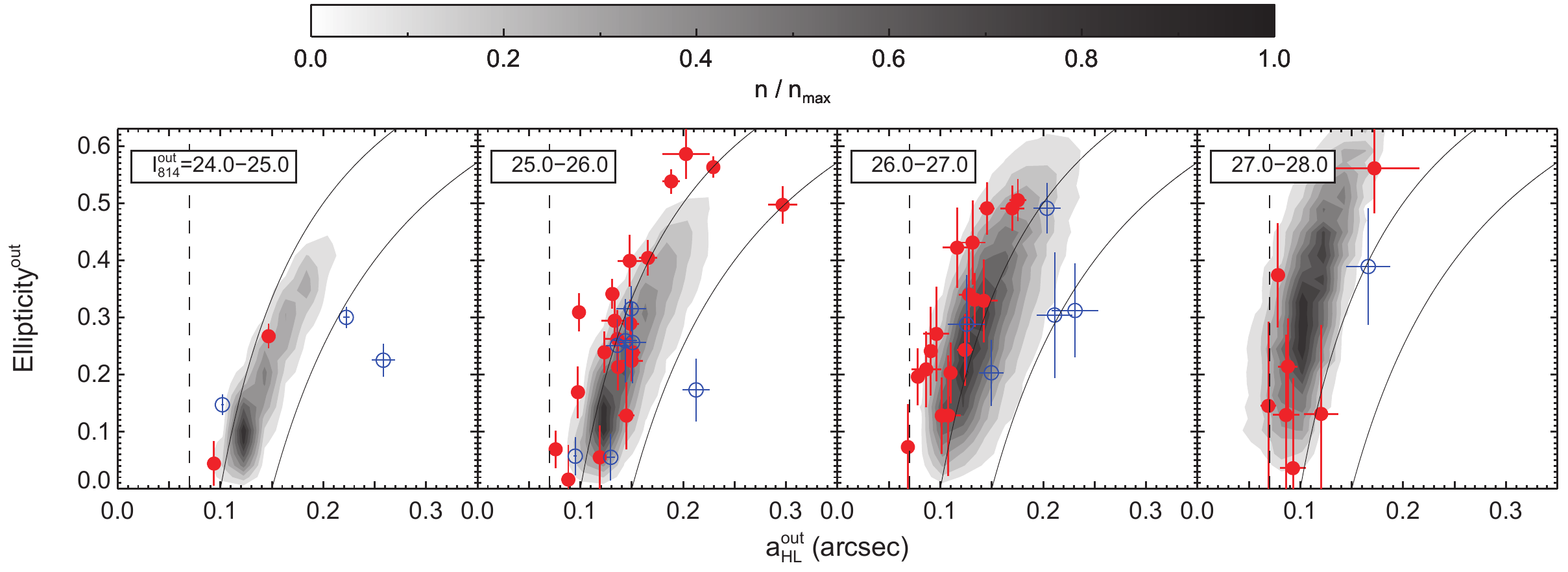}

    \caption{Same as Figure~\ref{fig:MCsingle}, but for the results of
    a Monte Carlo simulation in the case that a pair of two identical
    objects is detected as a single-blended source.  5000 sources
    which consists of two identical objects with
    $a_\mathrm{HL}^\mathrm{in} = 0\farcs 07$ and $\epsilon^\mathrm{in}
    = 0.0$ are generated in each set of input parameters, that is,
    $I_{814}^\mathrm{in} = 25.5$--28.0~mag ($\Delta
    I_{814}^\mathrm{in} = 0.5$~mag) and angular separation of 0\farcs
    03--0\farcs 30 ($\Delta r_\mathrm{sep} = 0\farcs 03$).  Note that
    the grayscales are evaluated only from the artificial sources
    detected as single sources (i.e., the sources detected as double
    detached sources are neglected).  The solid curves show the
    expected correlation between $\epsilon^\mathrm{out}$ and
    $a_\mathrm{HL}^\mathrm{out}$ through the separation between two
    identical objects $r_\mathrm{sep}$, which is represented as
    Equations~(\ref{eq-ExpectedCorr1}) and (\ref{eq-ExpectedCorr2}),
    with $a_\mathrm{HL}^\mathrm{in} = 0\farcs 10$ (left) and 0\farcs
    15 (right). \label{fig:MCdouble}}

   \end{figure*}
   Another possible origin of the positive correlation between
   $\epsilon$ and $a_\mathrm{HL}$ is \textit{blending} with unresolved
   double or multiple sources.  Let us consider a simplified situation
   where two identical round-shaped objects with half-light radius of
   $a_\mathrm{HL}^\mathrm{in}$ are located closely with a separation
   of $r_\mathrm{sep}$.  If these objects are blended as a single
   elongated source because of their small angular separation (i.e.,
   $r_\mathrm{sep} \lesssim 2 a_\mathrm{HL}^\mathrm{in}$), its
   half-light major radius of $a_\mathrm{HL}^\mathrm{out}$ and
   ellipticity of $\epsilon^\mathrm{out}$ can be roughly parameterized
   with $a_\mathrm{HL}^\mathrm{in}$ and $r_\mathrm{sep}$ as
   \begin{eqnarray}
    a_\mathrm{HL}^\mathrm{out} & \approx & \left(r_\mathrm{sep} +
     2a_\mathrm{HL}^\mathrm{in}\right) / 2, \label{eq-ExpectedCorr1}\\
    \epsilon^\mathrm{out} & \approx &
     r_\mathrm{sep} / \left(r_\mathrm{sep} +
		       2a_\mathrm{HL}^\mathrm{in} \right).
     \label{eq-ExpectedCorr2}
   \end{eqnarray}
   Since both $a_\mathrm{HL}^\mathrm{out}$ and $\epsilon^\mathrm{out}$
   are found to increase with $r_\mathrm{sep}$, a positive correlation
   between these two quantities will emerge even though each component
   has a perfectly round shape.  Under this interpretation, the
   observed correlation contains useful information that the LAEs may
   consist of two or more components with small angular separation.
   In this case, the correlation can be considered as an intrinsic one
   not an apparent one described in the previous subsection.

   In order to examine whether this interpretation results in a
   similar distribution in the $\epsilon$--$a_\mathrm{HL}$ plane to
   the observed one quantitatively, we perform Monte Carlo simulations
   whose details are described in Section~\ref{subsec:MC_double}.  The
   resultant distribution of the artificial sources in the
   $\epsilon^\mathrm{out}$--$a_\mathrm{HL}^\mathrm{out}$ plane are
   shown in Figure~\ref{fig:MCdouble}.  Since the distributions for
   the sources with $I_{814}^\mathrm{out} > 26$~mag are completely
   determined by the effects of shot noise, the distributions for the
   double-component sources with $I_{814}^\mathrm{out} > 26$~mag are
   similar to those for the single-component sources with
   $I_{814}^\mathrm{out} > 26$~mag shown in Figures~\ref{fig:MCsingle}
   and \ref{fig:MCsingle_all}.  On the other hand, the distributions
   for the double-component sources with $I_{814}^\mathrm{out} <
   26$~mag are different from those for the single-component sources.
   And these distributions appear to be consistent with the expected
   distributions from the simplified situation shown by the solid
   curves in Figure~\ref{fig:MCdouble}.  While the observed
   distribution of the LAEs with $I_{814} < 26$~mag is reproduced
   well, the LAEs with relatively bright (i.e., $I_{814} \approx
   25$--26~mag) and large sizes and ellipticities (i.e.,
   $a_\mathrm{HL} \approx 0\farcs 20$ and $\epsilon \approx 0.45$) are
   failed to be reproduced.  This is because $\epsilon^\mathrm{out}$
   cannot be much larger than 0.5 via such a blending of two identical
   sources since the angular separation should be smaller than $\sim
   2a_\mathrm{HL}^\mathrm{in}$ in order to be detected as a
   single-blended source.  However, these LAEs may also be reproduced
   if non-zero intrinsic ellipticities are adopted.  Moreover, the
   observed distribution of the LAEs are reproduced even better if the
   artificial sources with $\epsilon^\mathrm{in} = 0$ have a Gaussian
   distribution peaked at $(a_\mathrm{HL}^\mathrm{in},\
   r_\mathrm{sep}) \sim (0\farcs 10,\ 0\farcs 15)$.

   We note that, in our simulation, only the artificial sources with
   large separation (i.e., $\gtrsim 0\farcs 3$) are well resolved into
   two detached sources.  This result explain the observed results
   that the double-component LAEs have angular separation of
   $r_\mathrm{sep} > 0\farcs 36$ as shown in
   Table~\ref{tab:EachDouble} and that the single-component LAEs have
   $a_\mathrm{HL} < 0\farcs 3$.  In this interpretation, some of the
   46 single-component LAEs may contain double or multiple components
   with close angular separations.  Moreover, some of the components
   in the 8 double-component LAEs can be further resolved into compact
   components; that is, they can be regarded as multiple-component
   LAEs which consists of three or more components.  Therefore, the
   double-component fraction in our sample could be as high as
   $\approx 100$\%.

  \subsection{Dependence of Ly$\alpha$ Line EW and Luminosity on
  Size}\label{subsec:EW+LLya-Size}
  \begin{figure}
   \plotone{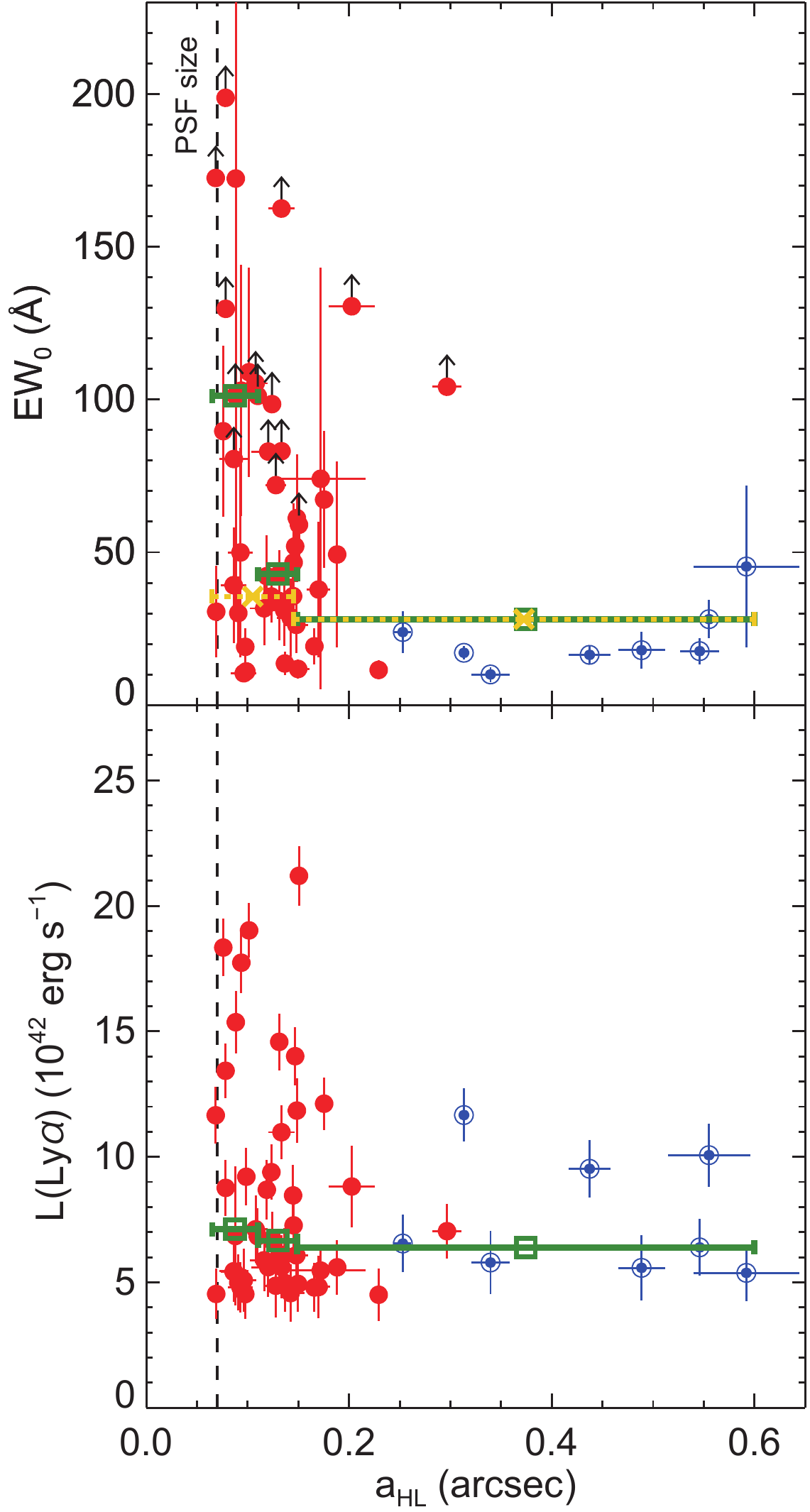}

   \caption{Distribution in the $\mathrm{EW_0}$--$a_\mathrm{HL}$ (top)
   and $L (\mathrm{Ly\alpha})$--$a_\mathrm{HL}$ planes (bottom).  The
   filled circles with upward arrows in the top panel indicate the
   lower limit of $\mathrm{EW_0}$.  The vertical dashed line indicates
   the PSF size of ACS image of 0\farcs 11.  The boxes with error bars
   represent the median values of $\mathrm{EW_0}$ (top) and $L
   (\mathrm{Ly\alpha})$ (bottom) for the ACS counterparts in each bin
   of $a_\mathrm{HL}$, in which the same number of the counterparts
   enters.  For the LAEs with lower limit of $\mathrm{EW_0}$, we use
   their $\mathrm{EW_0}$ lower-limit values to evaluate the median
   values.  On the other hand, the crosses with error bars in the top
   panel show the median values of $\mathrm{EW_0}$ for the case that
   the LAEs with lower limit of $\mathrm{EW_0}$ are neglected.
   \label{fig:EW+LLya-aHL}}

  \end{figure}
  It has been reported observationally that the high-$z$ LAEs and LBGs
  exhibit anti-correlation between size measured in rest-frame UV
  continuum and rest-frame Ly$\alpha$ EW, that is, the galaxies with
  large $\mathrm{EW_0}$ tend to have smaller sizes (e.g., Law et
  al. 2012; Vanzella et al. 2009; Pentericci et al. 2010; Shibuya et
  al. 2014; see also Bond et al. 2012 against these results).  As
  shown in the top panel of Figure~\ref{fig:EW+LLya-aHL}, our LAE
  sample at $z = 4.86$ also present such anti-correlation between
  $\mathrm{EW_0}$ and $a_\mathrm{HL}$.  However, this result seems to
  depend on the treatment of the LAEs with lower limits of
  $\mathrm{EW_0}$.  If they are included to calculate the
  binned-median values of $\mathrm{EW_0}$ using their lower limits of
  $\mathrm{EW_0}$, the anti-correlation between $\mathrm{EW_0}$ and
  $a_\mathrm{HL}$ is clearly seen as shown by the boxes with
  error-bars in the top panel of Figure~\ref{fig:EW+LLya-aHL}.  On the
  other hand, if they are completely neglected to calculate the
  robustly determined binned-median values, the anti-correlation
  disappears.  Therefore, in order to conclude whether or not the
  anti-correlation between $\mathrm{EW_0}$ and $a_\mathrm{HL}$ does
  exist, a deeper imaging data of the broadband which is used to
  determine $\mathrm{EW_0}$ (i.e., Subaru $z^\prime$ band for our LAE
  sample) is required.  Our LAE sample does not show strong
  correlation between $L (\mathrm{Ly\alpha})$ and $a_\mathrm{HL}$ as
  shown in the bottom panel of Figure~\ref{fig:EW+LLya-aHL}, while the
  dynamic range of $L (\mathrm{Ly\alpha})$ is only a factor of $\sim
  5$ and the maximum $L (\mathrm{Ly\alpha})$ seems to decrease as
  $a_\mathrm{HL}$ increases.

  Shibuya et al. (2014) found that, for their sample of the LAEs at $z
  \sim 2.2$, merger fraction decreases at large $\mathrm{EW_0}$.  If
  we consider the double-component LAEs as merging galaxies, as
  discussed in Section~\ref{subsec:Implication}, the same trend is
  also seen in our sample as shown in Figure~\ref{fig:EW+LLya-aHL};
  all double-component LAEs have $\mathrm{EW_0} < 50$~{\AA}.  Same
  trend hold for $L(\mathrm{Ly\alpha})$.  However, as presented in
  Section~\ref{subsec:dis-mor}, we cannot rule out the possibility
  that the single-component LAEs are merging galaxies.  We note that,
  although the trend seen in the $\mathrm{EW_0}$--$a_\mathrm{HL}$
  plane has been usually interpreted as the absence of the galaxies
  with large stellar mass (i.e., large in size) and large
  $\mathrm{EW_0}$, the trend is consistent with the model in which the
  galaxy merger and/or close encounter will activate Ly$\alpha$
  emission.  This is because the single-component LAEs can contain the
  galaxies with much smaller separations than the double-component
  LAEs and because galaxy pairs with smaller separations can result in
  more enhanced star formation as found in the nearby universe using
  the Sloan Digital Sky Survey (Patton et al. 2013).  Moreover, based
  on this scenario, since the single-component LAEs can contain both
  of the galaxies with short and long elapsed times from galaxy
  merger/interaction which activates Ly$\alpha$ emission, the median
  values of $\mathrm{EW_0}$ and $L (\mathrm{Ly\alpha})$ may not depend
  on the separation.  This expectation is also consistent with the
  observed distributions of the LAEs shown in
  Figure~\ref{fig:EW+LLya-aHL}.

  \subsection{Implication for Star Formation in the LAEs at $z =
  4.86$}\label{subsec:Implication}

  We detected 54 counterparts in the ACS images for our LAEs at $z =
  4.86$ in the COSMOS field.  While 8 of them have double component
  with the angular separations of 0\farcs 36--0\farcs 98 (i.e.,
  2.3--6.2~kpc at $z = 4.86$), the magnitudes and morphologies of
  individual components were found to be similar to those of the other
  46 single-component LAEs (see Figure~\ref{fig:e-aHL-sep} and
  Tables~\ref{tab:z4p9LAE} and \ref{tab:EachDouble}) and the typical
  LAEs in the literature (e.g., Malhotra et al. 2012; Hagen et
  al. 2014).  This result indicates that the double-component LAEs are
  interacting and/or merging galaxies with close separation, that is,
  the projected separation is comparable to or not larger than ten
  times of the size of a galaxy, $r_\mathrm{sep} / a_\mathrm{HL} \sim
  1$--10.

  Moreover, as shown in Section~\ref{subsec:dis-mor} through our Monte
  Carlo simulations, the observed positive correlation between
  $\epsilon$ and $a_\mathrm{HL}$ for our ACS-detected LAEs may
  indicate that both of the single-component LAEs and the individual
  components in the double-component LAEs consist of unresolved
  components with close separation of $r_\mathrm{sep} \lesssim 0\farcs
  3$ (i.e., $\lesssim 1.9$~kpc at $z = 4.86$), while another
  interpretation for the observed correlation (e.g., apparent
  correlation caused by the deformation effects such as PSF broadening
  and shot noise) was still possible.  Our Monte Carlo simulation also
  indicates that a typical size of individual component is $\sim
  0\farcs 10$--0\farcs 15 (i.e., $\sim 0.64$--0.96~kpc at $z = 4.86$).
  Since the observed wavelength of the ACS F814W-band corresponds to
  rest-frame UV wavelength of $\sim 1200$--1640~{\AA} at $z = 4.86$,
  the ACS components are considered to be young star-forming regions.
  Therefore, the small separation suggests the following two cases:
  (1) the individual component is a large star-forming region in an
  extended galaxy and star-formation activity in the LAEs occurs in a
  clumpy fashion or (2) individual component in an ACS source is a
  compact star-forming galaxy and the LAEs are the galaxies in close
  encounter and/or merger.  In order to distinct the above two
  interpretations, deeper imaging data at longer wavelength with
  similar or higher spatial-resolution than our ACS F814W-band data is
  inevitable.  If diffuse and faint underlying component which is
  surrounding the two (or multiple) components is detected and it does
  not show any signatures of galaxy interaction/merger, the clumpy
  star-formation in a galaxy will be confirmed.

  \begin{figure}
   \plotone{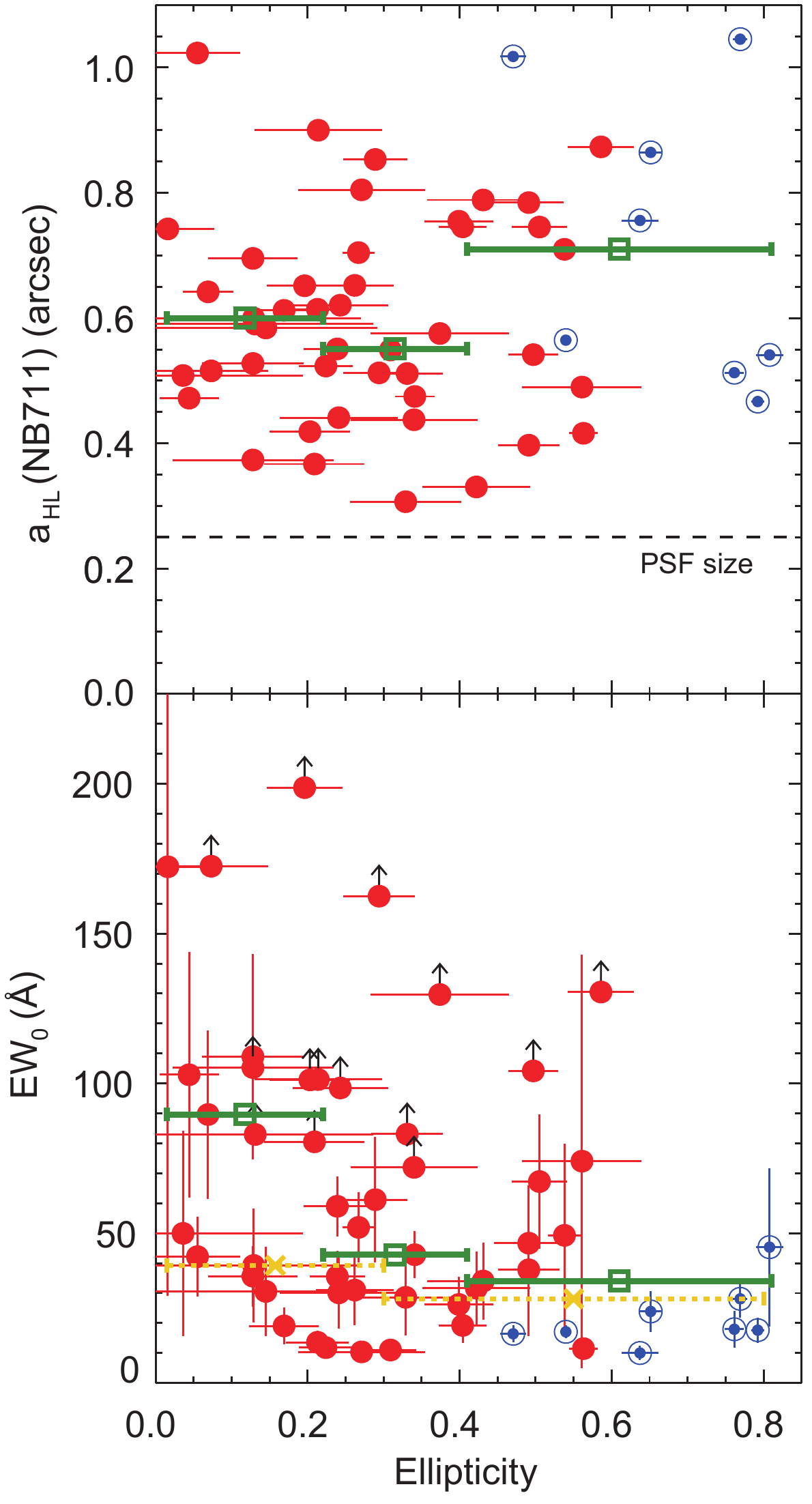}

   \caption{Same as Figure~\ref{fig:EW+LLya-aHL}, but for distribution
   in the $a_\mathrm{HL} (\mathrm{NB711})$--$\epsilon$ (top) and
   $\mathrm{EW_0}$--$\epsilon$ planes (bottom).  The horizontal dotted
   line in the top panel indicates the PSF size of NB711--band image
   of 0\farcs 25. \label{fig:Epsilon-FWHM}}

  \end{figure}
  In the interpretation of clumpy star-formation in a disk-like
  galaxy, as usually observed in high-$z$ galaxies (e.g., Elmegreen et
  al. 2009; F{\"o}rster Schreiber et al. 2011; Murata et al. 2014;
  Tadaki et al. 2014), the ellipticity of a source may be an intrinsic
  property related to the viewing angle of the disk.  That is, large
  ellipticity implies that its viewing angle is close to edge-on and
  that stellar disk lies in the elongated direction.  If we consider
  that Ly$\alpha$ is emitted in directions perpendicular to the disk,
  as predicted by the recent theoretical studies for Ly$\alpha$ line
  transfer (e.g., Verhamme et al. 2012; Yajima et al. 2012b), the
  pitch angle of Ly$\alpha$ emission will be at right angles to that
  of UV continuum.  Moreover, in such case, it is also expected that
  the size in Ly$\alpha$ emission $a_\mathrm{HL} (\mathrm{NB711})$
  shows a positive correlation with ellipticity measured in rest-frame
  UV continuum because Ly$\alpha$ emitting region in bipolar
  directions perpendicular to the disk can be viewed in longer
  distance if the viewing angle of the disk is closer to edge-on, that
  is, larger ellipticity.  However, as presented in the top panel of
  Figure~\ref{fig:Epsilon-FWHM}, we do not find such positive
  correlation between $a_\mathrm{HL} (\mathrm{NB711})$ and ellipticity
  for the 54 ACS-detected LAEs.  Furthermore, as shown in the bottom
  panel of Figure~\ref{fig:Epsilon-FWHM}\footnote{Note that,
  considering the strong positive correlation between $a_\mathrm{HL}$
  and $\epsilon$ shown in Figure~\ref{fig:e-aHL-total}, this plot is
  qualitatively identical to the distribution in the
  $\mathrm{EW_0}$--$a_\mathrm{HL}$ plane shown in the top panel of
  Figure~\ref{fig:EW+LLya-aHL}.}, the observed distribution of the
  LAEs in the $\mathrm{EW_0}$--$\epsilon$ plane seems not to be
  quantitatively consistent with the interpretation of clumpy
  star-formation in a disk-like galaxy, where $\mathrm{EW_0}$ is
  expected to decrease significantly toward edge-on direction (i.e.,
  larger ellipticity) via radiative transfer effects for Ly$\alpha$
  resonance photons (e.g., Verhamme et al. 2012; Yajima et al. 2012b);
  this result is consistent with Shibuya et al. (2014).  Therefore,
  the interpretation of clumpy star-formation in a disk-like galaxy
  seems not to be preferred for our LAE sample.  This conclusion can
  be reinforced with the absence of the ACS source with large size and
  round shape; if there are multiple clumpy star-forming regions in a
  disk-like galaxy, some of such galaxies will be viewed from face-on,
  resulting in large size and round shape.  We emphasize again that
  this result is not affected by a selection bias against them if they
  are bright enough (i.e., $I_{814} \lesssim 26$~mag) as shown in
  Figures~\ref{fig:size-I814} and \ref{fig:MC-detecomp}.

  On the other hand, the interpretation of merger and/or interaction
  is broadly consistent with these observed results.  The correlation
  between $\epsilon$ and $a_\mathrm{HL}$ can be reproduced by blending
  with double (or multiple) sources with close separations as shown in
  Section~\ref{subsubsec:DoubleDeform} through our Monte Carlo
  simulations.  The anti-correlation between the maximum value of
  $\mathrm{EW_0}$ or $L (\mathrm{Ly\alpha})$ and $a_\mathrm{HL}$ is
  also expected if the single-component LAEs are the merging and/or
  interacting galaxies with close separations and if Ly$\alpha$
  emissions are activated in such situation as described in
  Section~\ref{subsec:EW+LLya-Size}.  Moreover, the observed results
  that the median values of $\mathrm{EW_0}$ and $L
  (\mathrm{Ly\alpha})$ do not depend on $a_\mathrm{HL}$ are also
  consistent with this merger interpretation as shown in
  Section~\ref{subsec:EW+LLya-Size}.  Therefore, the interpretation of
  merging and/or interacting galaxies seems to be more feasible for
  our LAE samples.
 \section{CONCLUSIONS}

 We have examined the morphological properties of 61 LAEs at $z =
 4.86$ based on the \textit{HST}/ACS imaging in the F814W-band filter,
 which are originally selected in the COSMOS field by S09.  Our main
 results and conclusions are summarized below.

 \begin{enumerate}
  \item While the ACS counterparts of 7 LAEs are not detected, 62 ACS
	sources are detected with $I_{814} \lesssim 28$~mag for the
	remaining 54 LAEs.  Of the 54 LAEs with ACS sources, 8 LAEs
	have double ACS components and 46 LAEs have single component.
  \item For the double-component 8 LAEs, the angular separation
	between two components are found to be 0\farcs 36--0\farcs 98
	($= 2.3$--6.2~kpc at $z = 4.86$) with a mean separation of
	0\farcs 63 ($= 4.0$~kpc).  The angular separation is
	sufficiently large compared to the PSF size of ACS image,
	$R_\mathrm{PSF} = 0\farcs 07$, which is the reason why they
	are separately detected.
  \item Comparing ACS F814W-band magnitude $I_{814}$ with Suprime-Cam
	NB711--, $i^\prime$--, and $z^\prime$--band magnitudes, we find
	that the ACS F814W-band image probes rest-frame UV continuum
	rather than Ly$\alpha$ line (Figure~\ref{fig:I814vsNB711}).
	We observe the extent of star-forming regions in our LAE
	sample at $z = 4.86$ via the F814W-band filter.
  \item All of 62 ACS sources have small spatial sizes of
	$a_\mathrm{HL} \sim 0\farcs 07$--0\farcs 30 ($=
	0.45$--1.9~kpc) as shown in Figure~\ref{fig:e-aHL-sep}.  Their
	mean size is 0\farcs 14
	($= 0.89$~kpc), which is consistent
	with the previous measurements for the size in rest-frame UV
	continuum of the LAEs at $z \sim 2$--6 in the literatures.
  \item The measured ellipticities of the 62 ACS sources are widely
	distributed in $\epsilon = 0.02$--0.59 and a positive
	correlation between $\epsilon$ and $a_\mathrm{HL}$
	(Figure~\ref{fig:e-aHL-sep}).  It is evident even if we
	exclude the faint ACS sources with $I_{814} > 26$~mag.
	Moreover, the absence of the large (i.e., $a_\mathrm{HL}
	\gtrsim 0\farcs 2$) sources with almost round shape (i.e.,
	$\epsilon \lesssim 0.2$) is also found.
  \item The 7 ACS-undetected LAEs are expected to have low surface
	brightnesses so that they are undetected in our ACS images.
	We estimate their half-light radii from Suprime-Cam
	$i^\prime$--band magnitudes of $i^\prime = 25.23$--27.04~mag
	(Figure~\ref{fig:ipDist}) to be $R_\mathrm{HL} \gtrsim 0\farcs
	07$--0\farcs 32.
  \item All ACS sources have significantly smaller sizes in UV
	continuum than those in Ly$\alpha$ lines probed by NB711--band
	(Figure~\ref{fig:RHLvsFWHM}).  The size ratios of
	$a_\mathrm{HL} (\mathrm{NB711}) / a_\mathrm{HL}
	(\mathrm{F814W})$ are widely distributed in the range of
	$\approx 1$--10.
  \item The observed positive correlation between $\epsilon$ and
	$a_\mathrm{HL}$ can be interpreted by either (1) an apparent
	one caused by the deformation effects such as the PSF
	broadening and shot noise or (2) an intrinsic one originated
	from blending with unresolved double or multiple sources.
	These are proved through our Monte Carlo simulations, which
	reproduce the observed correlations as presented in
	Figures~\ref{fig:MCsingle_all} and \ref{fig:MCdouble} for the
	former and latter interpretations, respectively.
  \item Both Ly$\alpha$ EW and luminosity of LAEs do not show strong
	dependencies on sizes in rest-frame UV continuum
	(Figure~\ref{fig:EW+LLya-aHL}).  Moreover, there are no LAEs
	with double ACS components at large $\mathrm{EW_0}$ and
	$L(\mathrm{Ly\alpha})$.  These results are consistent with the
	model in which galaxy merger and/or close encounter will
	activate Ly$\alpha$ emissions.
  \item The 8 double-component LAEs are considered to be merger and/or
	interacting galaxies since the angular separations between
	components are significantly larger than the sizes of each
	component although we cannot completely reject the possibility
	that their underlying (disk) component is missed by its
	faintness and they are single object with multiple
	star-forming knot.  The absence of the ACS sources with large
	sizes and small ellipticities (Figures~\ref{fig:e-aHL-total}
	and \ref{fig:e-aHL-sep}), the anti-correlation between
	$\mathrm{EW_0}$ or $L (\mathrm{Ly\alpha})$ and $a_\mathrm{HL}$
	(Figure~\ref{fig:EW+LLya-aHL}), and the absence of the
	correlation between $\epsilon$ and $a_\mathrm{HL}
	(\mathrm{NB711})$ (Figure~\ref{fig:Epsilon-FWHM}) suggest the
	possibility that a significant fraction of 46 single-component
	LAEs are also merger/interacting galaxies with a very small
	separation.  In order to decipher which interpretation is
	adequate for our LAE sample, further observation with high
	angular resolution at the wavelengths which are longer than
	the Balmer/4000~{\AA} break in rest frame (i.e., $\gtrsim
	2.3~\mu$m in observer frame for our LAE sample at $z = 4.86$)
	will be required.
 \end{enumerate}

\acknowledgements

We would like to thank both the Subaru and HST staff for their
invaluable help, all members of the COSMOS team, Tsutomu T. Takeuchi
at Nagoya university for his help in running our Monte Carlo
simulations using his computers, and Shinki Oyabu at Nagoya university
for providing valuable suggestions/comments.  We would also like to
thank the anonymous referees for his/her useful comments.  This work
is based on observations taken by the CANDELS Multi-Cycle Treasury
Program with the NASA/ESA HST, which is operated by the Association of
Universities for Research in Astronomy, Inc., under NASA contract
NAS5-26555.  This work was in part financially supported by JSPS
(15340059 and 17253001).


\appendix

 \section{Monte Carlo Simulation}

 In this Appendix, we describe the details of the settings and
 procedures of our Monte Carlo simulations.

  \subsection{Single Component}\label{subsec:MC_single}

  In order to estimate the detection completeness
  (Section~\ref{subsec:size}) and the deformation effects in shape via
  the PSF broadening and shot noise for faint single sources
  (Section~\ref{subsubsec:SingleDeform}), we performed the following
  Monte Carlo simulations.  For each artificial ACS F814W-band source,
  the exponential light profile is adopted, motivated by the
  observational result that the 47 ACS-detected LAEs at $z = 5.7$ in
  the COSMOS field have the S\'ersic index of $n = 0.7 \pm 0.3$
  (Taniguchi et al. 2009).  For each set of the given input parameters
  of $I_{814}^\mathrm{in}$, $a_\mathrm{HL}^\mathrm{in}$, and
  $\epsilon^\mathrm{in}$, we prepare 1000 artificial sources by using
  the GALFIT software (Peng et al. 2002, 2010).  The full ranges
  (steps) of these input parameters are $I_{814}^\mathrm{in} =
  24.0$--28.0~mag ($\Delta I_{814}^\mathrm{in} = 0.5$~mag),
  $a_\mathrm{HL}^\mathrm{in} = 0\farcs 03$--0\farcs 75 ($\Delta
  a_\mathrm{HL}^\mathrm{in} = 0\farcs 03$), and $\epsilon^\mathrm{in}
  = 0.0$--0.9 ($\Delta \epsilon^\mathrm{in} = 0.1$); in total, we
  prepare 2,250,000 artificial sources.  We put them into the observed
  ACS image randomly, convolving them with the PSF image.  The photon
  noises and the Galactic dust extinction (i.e., $A_\mathrm{814W} =
  0.035$; Capak et al. 2007) are also added to them.  Then we detect
  sources and measure their photometric properties, that is, ACS
  F814W-band magnitude $I_{814}^\mathrm{out}$, half-light major radius
  $a_\mathrm{HL}^\mathrm{out}$, and ellipticity
  $\epsilon^\mathrm{out}$, with the same procedure to the observed ACS
  images using the SExtractor modified by one of the authors as
  described in Section~\ref{sec:data}.

  \begin{figure*}
   \plotone{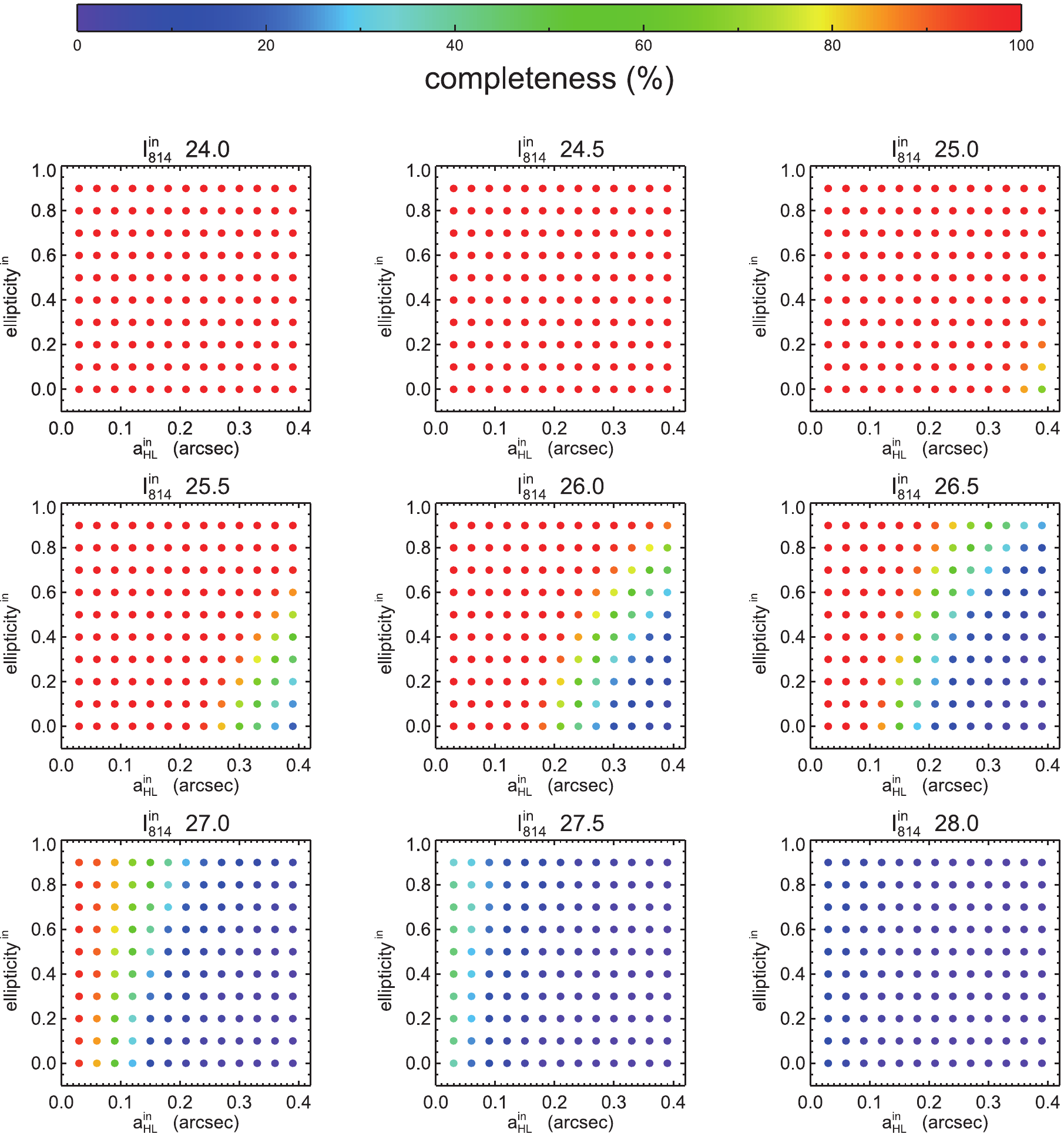}
   \caption{2 dimensional map of the detection completeness in the
   $\epsilon^\mathrm{in}$--$a_\mathrm{HL}^\mathrm{in}$ plane estimated
   through a Monte Carlo simulation.  The map is separately plotted
   for different input magnitude of
   $I_{814}^\mathrm{in}$. \label{fig:MC-detecomp}}
  \end{figure*}
  The resultant detection completenesses as a function of the input
  parameters of $I_{814}^\mathrm{in}$, $a_\mathrm{HL}^\mathrm{in}$,
  and $\epsilon^\mathrm{in}$ are presented in
  Figure~\ref{fig:MC-detecomp}.  As expected, the sources with
  relatively bright input magnitudes of $I_{814}^\mathrm{in} \lesssim
  25.0$~mag have high detection completenesses of $\approx 100$\%
  regardless of the other input parameters of
  $a_\mathrm{HL}^\mathrm{in}$ and $\epsilon^\mathrm{in}$.  For the
  fainter sources with $I_{814} \gtrsim 25.0$~mag, the detection
  completenesses are found to become lower for more extended sources
  with smaller ellipticities.  This is simply because, at a given
  $I_{814}^\mathrm{in}$, the surface brightnesses are lower for the
  sources with large $a_\mathrm{HL}^\mathrm{in}$ at a fixed
  $\epsilon^\mathrm{in}$ and for those with small
  $\epsilon^\mathrm{in}$ at a fixed $a_\mathrm{HL}^\mathrm{in}$.  For
  the sources fainter than the $3\sigma$ limiting magnitude of the ACS
  images in a $1^{\prime\prime}$ diameter aperture of 27.4~mag, almost
  all sources are found to be undetected regardless of
  $a_\mathrm{HL}^\mathrm{in}$ and $\epsilon^\mathrm{in}$.

  \subsection{Double Component}\label{subsec:MC_double}

  Similar to the Monte Carlo simulations for single-component sources
  described in Section~\ref{subsec:MC_single}, we performed the
  following simulations in order to estimate the deformation effects
  in shape via the blending with unresolved double or multiple sources
  (Section~\ref{subsubsec:DoubleDeform}).  For simplicity, all
  artificial sources are assumed to consist of two identical
  components having the same $a_\mathrm{HL}^\mathrm{in}$ and
  $I_{814}^\mathrm{in}$ with angular separation of $r_\mathrm{sep}$.
  Each component is assumed to have a perfectly round shape (i.e.,
  $\epsilon^\mathrm{in} = 0$).  As done in
  Section~\ref{subsec:MC_single}, the exponential light profile is
  adopted for each component.  We adopt a single value of 0\farcs 07
  for the intrinsic size of each component of
  $a_\mathrm{HL}^\mathrm{in}$, which is found to result in a similar
  distribution in the $\epsilon$--$a_\mathrm{HL}$ plane to the
  observed one.  The full ranges (steps) of the other parameters are
  provided as $I_{814}^\mathrm{in} = 25.5$--28.0~mag ($\Delta
  I_{814}^\mathrm{in} = 0.5$~mag) and $r_\mathrm{sep} = 0\farcs
  03$--0\farcs 30 ($\Delta r_\mathrm{sep} = 0\farcs 03$).  We prepare
  5000 sources for each set of parameters; in total, 300,000
  artificial sources are generated.  Then, as done in
  Section~\ref{subsec:MC_single}, we put these sources on the observed
  ACS image randomly, convolving the PSF image and adding photon
  noises and the Galactic dust extinction.  We try to extract their
  images by using SExtractor with the same parameter set described in
  Section~\ref{sec:data}.  The resultant magnitude
  $I_{814}^\mathrm{out}$, half-light major radius
  $a_\mathrm{HL}^\mathrm{out}$, and ellipticity
  $\epsilon^\mathrm{out}$ are measured only for the sources detected
  as single component using the SExtractor modified by one of the
  authors.
\end{document}